\documentclass[aps,preprint,showpacs,nofootinbib,showkeys,tightenlines]{revtex4}
\usepackage{graphicx}  %
\usepackage{amsmath}
\usepackage{bm}  %
\usepackage{ulem}  %
\usepackage{verbatim} %

\newcommand{\bea}{\begin{eqnarray}}
\newcommand{\eea}{\end{eqnarray}}
\newcommand{\beq}{\begin{equation}}
\newcommand{\eeq}{\end{equation}}
\newcommand{\bqa}{\begin{eqnarray}}
\newcommand{\eqa}{\end{eqnarray}}

\def\mqo2{{\!\!\!}}

\usepackage{relsize}
\def\babar{\mbox{\slshape B\kern-0.1em{\smaller A}\kern-0.1em
    B\kern-0.1em{\smaller A\kern-0.2em R}}}

\begin{document}

\title{
Masses of Doubly Heavy Tetraquarks  with Error Bars
}
\author{Eric Braaten}
%\email{braaten@mps.ohio-state.edu}
\affiliation{Department of Physics,
         The Ohio State University, Columbus, OH\ 43210, USA}
\author{Li-Ping He}
\affiliation{Department of Physics,
         The Ohio State University, Columbus, OH\ 43210, USA}
\author{Abhishek Mohapatra}
\affiliation{Department of Physics, Duke University, Durham, NC 27705, USA}
\date{\today}
%\date{November 2007}

\begin{abstract}
In the heavy-quark limit,
the two heavy quarks in a doubly heavy baryon or a doubly heavy tetraquark
are bound by their color-Coulomb potential into a compact diquark.
The doubly heavy hadrons are related by the approximate heavy-quark--diquark symmetry of QCD
 to the heavy hadrons obtained by replacing the heavy diquark  by a heavy antiquark.
Effective field theories can be used to expand the masses of singly heavy hadrons and doubly heavy hadrons
in inverse powers of the heavy quark masses.
The coefficients in the expansions for doubly heavy tetraquarks can be determined from those for 
heavy mesons, heavy baryons, and doubly heavy baryons using heavy-quark--diquark symmetry.
We predict the masses of the 
ground-state doubly heavy tetraquarks with error bars
using as inputs  the masses of heavy mesons and heavy baryons measured in experiments 
and the masses of doubly heavy baryons calculated using lattice QCD.
The only doubly heavy tetraquarks predicted to be stable with respect to strong decays are 
$bb$ tetraquarks with light flavor $\bar u \bar d$,  $\bar s \bar u$ and $\bar s \bar d$.
 \end{abstract}

\smallskip
\pacs{14.20.Lq, 14.20.Mr, 14.40.Rt}
\keywords{
doubly heavy tetraquarks, doubly heavy baryons, heavy-diquark limit.}
\maketitle

%%%%%%%%%%%%%%%%%%%%%%%%%%%%%%%%%%%%%%%%%%%%%%%
\section{Introduction}
\label{sec:intro}
%%%%%%%%%%%%%%%%%%%%%%%%%%%%%%%%%%%%%%%%%%%%%%%

One of the most basic properties of a quantum field theory is its particle spectrum.
The stable particles are particularly important, because they can exist as asymptotic states.
The spectrum of quantum chromodynamics (QCD) consists of color-singlet clusters of quarks and gluons,
including 2-quark ($q \bar q$) mesons and 3-quark ($qqq$) baryons.
The flavor symmetries of QCD guarantee that the lightest 2-quark meson with any net flavor 
and the lightest 3-quark baryon with any flavor are stable with respect to QCD interactions, so they can decay only through electromagnetic or weak interactions.
The spectrum of QCD may also include exotic hadrons with additional constituents,
such as tetraquark ($qq \bar q \bar q$) mesons and pentaquark ($qqqq \bar q$) baryons.
Since the constituents of such an exotic hadron can be rearranged into two color-singlet clusters,
it can decay into two hadrons unless the mass of the exotic hadron is below the strong-decay threshold.
Whether there are any such exotic hadrons that are stable is a dynamical question that depends on
the parameters of QCD and specifically on the quark masses.

The possibility that the spectrum of QCD includes stable tetraquark mesons 
containing two heavy quarks ($QQ \bar q \bar q$) was first studied by 
Richard and collaborators using quark potential models \cite{Ader:1981db,Zouzou:1986qh}.
Manohar and Wise pointed out that QCD predicts that there must be stable
$QQ \bar q \bar q$ tetraquark mesons
in the  limit of  infinite heavy-quark mass \cite{Manohar:1992nd}.
In this limit, the attractive color-Coulomb potential between the two heavy quarks 
decreases the energy of the $Q Q \bar q \bar q$ meson to below the energy of two separated $Q\bar q$ mesons.
The color-Coulomb potential binds the $QQ$ pair 
into a compact diquark whose effect on the lighter QCD fields is
the same as that of a single heavy antiquark. 
The resulting approximate symmetry of QCD, which is called heavy-quark--diquark symmetry,
relates doubly heavy $QQq$ baryons to heavy $\bar Q q$ mesons
\cite{Savage:1990di,Brambilla:2005yk,Fleming:2005pd} 
and  doubly heavy $QQ \bar q\bar q$ tetraquarks to heavy $\bar Q \bar q  \bar q$ 
anti-baryons \cite{Mehen:2017nrh}.

Whether QCD predicts the existence of stable doubly heavy tetraquarks containing bottom or charm quarks
is a dynamical question that can only be answered using quantitative information from QCD.
Bicudo  {\it et al.}\ used lattice QCD calculations of static potentials 
for two heavy quarks together with the Born-Oppenheimer approximation to present evidence for the existence 
of a stable tetraquark with flavor $bb\bar u \bar d$ 
\cite{Bicudo:2015vta,Bicudo:2016ooe}.
A convincing case that QCD predicts the existence of a stable  $bb\bar u \bar d$ tetraquark
was made by Karliner and Rosner \cite{Karliner:2017qjm} and by Eichten and Quigg \cite{Eichten:2017ffp}.
Karliner and Rosner also predicted that there are $bc$ and $cc$ tetraquarks 
with masses near their  strong-decay thresholds.
Eichten and Quigg also predicted that there are  stable 
$bb\bar s \bar u$ and $bb\bar s \bar d$ tetraquarks,
but that all $bc$ and $cc$ tetraquarks have masses well above their strong-decay thresholds.
Doubly heavy tetraquarks have also been studied extensively using quark potential models and diquark models. A fairly comprehensive list of references is given in Ref.~\cite{Lu:2020rog}.

Doubly heavy tetraquarks have been investigated directly using lattice QCD.
Francis {\it et al.}\ presented strong evidence for the existence of deeply bound tetraquarks 
with flavors $bb\bar u \bar d$,  $bb\bar s \bar u$, and $bb\bar s \bar d$ \cite{Francis:2016hui,Francis:2018jyb}. 
The Hadron Spectrum Collaboration studied $cc$ tetraquarks in lattice QCD  
with unphysically large $u$ and $d$ masses and found no evidence for bound states \cite{Cheung:2017tnt}.
Junnarkar, Mathur and Padmanath verified the existence of deeply bound tetraquarks 
with flavors $bb\bar u \bar d$,  $bb\bar s \bar u$, and $bb\bar s \bar d$  \cite{Junnarkar:2018twb}.
Leskovec {\it et al.}\ calculated the mass of the ground-state $bb\bar u \bar d$ tetraquark  with quantum numbers $1^+$
with all the major systematic errors quantified  \cite{Leskovec:2019ioa}.

Lattice QCD should eventually be able to provide definitive predictions for the masses
of all the doubly heavy tetraquarks.  However other methods based on QCD can still provide useful
insights into the spectrum of doubly heavy tetraquarks, especially if the errors in their predictions can be quantified.
In this paper, we quantify the errors in the approach used by Eichten and Quigg to demonstrate the existence of 
stable $bb$ tetraquarks \cite{Eichten:2017ffp}.  
The weakest point in their analysis was in their values for  the masses of doubly heavy baryons.  
To determine the  masses of $cc$ tetraquarks, they used the LHCb measurement of the mass of the 
double-charm baryon $\Xi_{cc}^{++}$, which has a well defined error bar \cite{Aaij:2017ueg}. 
However to determine the masses of  $bc$ and $bb$ tetraquarks,
they used predictions for the masses of $bc$ and $bb$ baryons from a quark model \cite{Karliner:2014gca}. 
We instead use the masses of doubly heavy baryons calculated using lattice QCD, which have well-defined error bars.
This allows us to give predictions for the masses of  doubly heavy tetraquarks in the heavy-diquark limit with error bars.
We also improve on the analysis of Ref.~\cite{Eichten:2017ffp} in several other ways,
including correcting an error in the heavy-quark mass dependence of 
the coefficient in an effective Hamiltonian for doubly heavy hadrons.

In Section~\ref{sec:1Heavy},  we use the measured masses of 
the ground-state heavy mesons and the ground-state heavy baryons
to determine coefficients in the expansions of their masses in inverse powers of the heavy-quark mass. 
We also use the masses of ground-state heavy baryons calculated using lattice QCD
to determine coefficients in the expansions. 
An effective field theory for doubly heavy hadrons  \cite{Brambilla:2005yk,Fleming:2005pd} 
can be used to organize their masses into expansions in inverse powers of the heavy quark masses.
In Section~\ref{sec:2Heavycore}, we use the masses of  ground-state doubly heavy baryons 
calculated using lattice QCD to determine coefficients in the expansions. 
We then use heavy-quark--diquark symmetry to 
determine coefficients in the expansions of the masses for doubly heavy tetraquarks.
Finally we use these expansions to
predict the masses of  doubly heavy tetraquarks with error bars.
The only doubly heavy tetraquarks with masses below the strong-decay thresholds 
have flavors $bb\bar u \bar d$, $bb\bar s \bar u$, and $bb\bar s \bar d$.
We conclude in Section~\ref{sec:Discussion} by summarizing our results and
discussing the prospects for 
more accurate predictions of the masses for doubly heavy tetraquarks.

%\newpage

%%%%%%%%%%%%%%%%%%%%%%%%%%%%%%%%%%%%%%%%%%%%%%%
\section{Singly Heavy Hadrons}
\label{sec:1Heavy}
%%%%%%%%%%%%%%%%%%%%%%%%%%%%%%%%%%%%%%%%%%%%%%%

In this section, we consider mesons and baryons that contain a single heavy quark.
Heavy Quark Effective Theory (HQET) can be used to organize their masses into expansions 
in inverse powers of the heavy quark mass $m_Q$.
We determine coefficients in the expansions of the masses of heavy mesons and heavy baryons 
using the masses measured in experiments.
We also determine  coefficients in the expansions of the masses of heavy baryons 
using masses calculated using lattice QCD.

%%%%%%%%%%%%%%%%%%%%%%%%%%%%%%%%%%%%%%%%%%%%%%%
\subsection{Heavy Mesons}
\label{sec:HeavyMeson}

A heavy hadron contains a single heavy quark $Q$ and light QCD fields.
The flavor  of the heavy quark  can be charm ($c$) or bottom ($b$).  
It has spin quantum number $\tfrac12$, and it is in a color-triplet ({\bf 3}) state.
In the presence of a heavy quark $Q$, the light QCD fields can have finite energy 
only if they can combine with the heavy quark
to form a color singlet.  This constrains the flavor of the light QCD fields.
The simplest possibility is the flavor of a light antiquark $\bar q$.
The corresponding hadron is a heavy meson with flavor $Q\bar q$.
In the presence of the heavy quark, the lowest-energy states of the light QCD fields are discrete states $\ell$
that can be specified by several quantum numbers:
\begin{itemize}
\item
three possible flavors $\bar q$.
They can be arranged into an isospin doublet $(\bar u,\bar d)$ and an isospin singlet $\bar s$.
\item
the angular-momentum/parity quantum numbers $j^P$, where $j =\tfrac12, \tfrac32,\tfrac52, \ldots$
is a half-odd-integer and $P =+,-$.
\item
a principal quantum number $\nu =1, 2, \ldots$  that labels successive states with the same $j^P$. 
\end{itemize}
In a heavy meson, the light QCD states with the lowest principal quantum number can  be labeled by $\ell = \bar q, j^P$. 

HQET is an effective field theory for the sector of QCD 
with  a single heavy hadron  \cite{Georgi:1990um}. 
The Lagrangian for HQET is organized  
into an expansion in powers of the inverse heavy quark mass $1/m_Q$.
At first order in $1/m_Q$, the terms in the Lagrangian 
are a kinetic-energy term and a spin-dependent term that depends on the spin $\bm{S}$ of the heavy quark.
From the form of the HQET Lagrangian, we can infer the form of the Hamiltonian $H^Q_\ell$
for a heavy meson with light QCD fields in the state $\ell$.
Through first order in $1/m_Q$, the only operators in the Hamiltonian are the heavy-quark spin $\bm{S}$ and the total 
angular momentum $\bm{j}_\ell$ of the light QCD fields. The quantum number for  $\bm{j}_\ell$ is the half-odd integer $j$.
The Hamiltonian $H^Q_\ell$ 
for a heavy meson through first order in $1/m_Q$ is
%%%%%%%%
\begin{equation}
H^Q_\ell = m_Q + \mathcal{E}_\ell + \frac{\mathcal{K}_\ell}{2 m_Q} 
+ \frac{\mathcal{S}_\ell}{2 m_Q} \bm{S} \cdot \bm{j}_\ell.
\label{H_ell}
\end{equation}
%%%%%%%%
The only dependence on the heavy flavor $Q$ is through the mass $m_Q$.
The coefficients  $\mathcal{E}_\ell$, $\mathcal{K}_\ell$, and $\mathcal{S}_\ell$
depend on the state $\ell$ of the light QCD fields.
The total angular momentum of the heavy meson is $\bm{J} = \bm{S} + \bm{j}_\ell$.
The  quantum number for $\bm{J}$ is the spin  $J$ of the heavy meson.
The heavy mesons with light QCD state  $\ell = \bar q, j^P$ 
form a spin doublet consisting of two states with spins 
$J=j - \tfrac12,j+\tfrac12$ whose mass splitting is proportional to $1/m_Q$.
That spin splitting is referred to as a hyperfine splitting.

The lowest-energy states $\ell$ of the light QCD fields
can be deduced from the masses of the observed  heavy mesons containing a single charm or bottom quark.
The ground state  of the light QCD fields for a given light flavor $\bar q$ has $j^P = \tfrac12^-$. 
It can be interpreted as a constituent antiquark in an S-wave orbital.
The ground-state heavy meson with  light flavor $\bar q$  
is therefore a spin doublet consisting of two states with 
$J^P=0^-,1^-$.
The charm-meson doublets with light flavors $\bar u$, $\bar d$, and $\bar s$
are $(D^0, D^{*0})$, $(D^+, D^{*+})$, and $(D_s^+, D_s^{*+})$.
The bottom-meson doublets with light flavors $\bar u$, $\bar d$, and $\bar s$
are $(B^-, B^{*-})$, $(\bar B^0, \bar B^{*0})$, and $(\bar B_s^0, \bar B_s^{*0})$.

%%%%%%%%%%%%%%%%%%%%%%%%%%%%%%%%%%%%%%%%%%%%%%%
\subsection{Heavy Baryons}
\label{sec:HeavyBaryon}

The second simplest possibility for the flavor of the light QCD fields in the presence of a heavy quark $Q$
is the  flavor $q q'$ of two light quarks.
The corresponding hadron is a heavy baryon with flavor $Q q q'$.
In the presence of the heavy quark, the lowest-energy states of the light QCD fields are discrete states $\ell$
that can be specified by several quantum numbers:
\begin{itemize}
\item
nine possible flavors $q q'$.
The antisymmetric flavors are an isospin singlet  $[u\,d]$ and an isospin doublet  $\big( [d\, s],[u\, s] \big)$.
The symmetric flavors are an isospin triplet  $\big( dd,  \{u\, d \}, uu \big)$,
 an isospin doublet   $\big( \{d\,s \}, \{u\, s \} \big)$, and an isospin singlet  $ss$.
\item
the angular-momentum/parity quantum numbers $j^P$, where $j =0, 1,2, \ldots$ is an integer
and $P =+,-$,
\item
a principal quantum number  $\nu =1,2, \ldots$ that labels successive states with the same $j^P$.  
\end{itemize}
In a heavy baryon, the  light QCD states $\ell$ with  the lowest principal  quantum number
can be labeled by  $[q\, q^\prime], j^P$ or $\{q\, q^\prime \}, j^P$.

From the form of the HQET Lagrangian, we can infer the form of the Hamiltonian $H^Q_\ell$
for a heavy baryon with light QCD fields in the state $\ell$.
Through first order in $1/m_Q$, the only operators in the Hamiltonian are the heavy-quark spin $\bm{S}$ 
and the total angular momentum $\bm{j}_\ell$ of the light QCD fields.  
The quantum number for $\bm{j}_\ell$ is the integer $j$.
The Hamiltonian $H^Q_\ell$ for a heavy baryon 
through first order in $1/m_Q$ has the same form as Eq.~\eqref{H_ell}.
The total angular momentum of the heavy baryon is $\bm{J} = \bm{S} + \bm{j}_\ell$.
The quantum number for $\bm{J}$ is the spin $J$  of the heavy baryon.
If $j=0$, the heavy baryon is a spin singlet with spin $J=\tfrac12$.
If $j \ge 1$, the heavy baryons form a spin doublet consisting of 
two states with spins $J=j - \tfrac12,j+\tfrac12$
whose mass splitting is proportional to $1/m_Q$.
That spin splitting is referred to as a hyperfine splitting.

The lowest-energy states $\ell$ of the light QCD fields
can be deduced from the masses of the observed heavy baryons containing a single charm or bottom quark.
For an antisymmetric light flavor $[q\, q^\prime]$, 
the ground state  of the light QCD fields has $j^P = 0^+$.  
It can be interpreted as two constituent quarks that are antisymmetric in their colors, 
antisymmetric in their spins, and in S-wave orbitals.
The ground-state heavy baryon with antisymmetric light flavor  $[q\, q^\prime]$
is a spin singlet  with $J^P=\tfrac12^+$.
For heavy flavor $Q$, the baryon with the isospin-singlet light flavor $[u\, d]$  is  $\Lambda_Q$.
The baryons with the isospin-doublet  light flavors $([d\,s], [u\, s])$ are  $\Xi_Q$, 
with a superscript that specifies the electric charge.
For a symmetric light flavor $ \{q \, q^\prime \}$, 
the ground state  of the light QCD fields has $j^P = 1^+$.
It can be interpreted as two constituent quarks that are antisymmetric in their colors, 
symmetric in their spins, and in S-wave orbitals.
The ground-state heavy baryons with symmetric flavor  $ \{q \, q^\prime \}$
are spin doublets consisting of two states with 
$J^P=\tfrac12^+,\tfrac32^+$.
For a  heavy flavor $Q$, the spin doublets  with the  isospin-triplet light flavors $(dd,\{u \,d\}, uu)$
are $(\Sigma_Q, \Sigma_Q^*)$. 
The  spin doublets  with  the   isospin-doublet  light flavors $(\{d \,s\},\{u \,s\})$ are  $(\Xi'_Q,\Xi_Q^*)$.
The  spin doublet with the light flavor $ss$ is $(\Omega_Q, \Omega_Q^*)$. 
For each spin doublet  of baryons, the members of an isospin multiplet are distinguished
by giving their electric charge as a superscript.

%%%%%%%%%%%%%%%%%%%%%%%%%%%%%%%%%%%%%%%%%%%%%%%
\subsection{Heavy Quark Masses}
\label{sec:HeavyQuarkMasses}

In Ref.~\cite{Eichten:2017ffp}, Eichten and Quigg presented an updated analysis of the $1/m_Q$ expansion
for  the masses of ground-state heavy hadrons. 
Their prescription for the heavy quark mass $m_Q$ was half the mass of the 
most deeply bound quarkonium state with $J^{PC} = 1^{--}$,
which is $J/\psi$ for charmonium and $\Upsilon$ for bottomonium.
The resulting masses for the charm and bottom quarks  are $m_c = 1.55$~GeV and $m_b=4.73$~GeV.
This prescription is not ideal, because it is biased by the  large binding energy of $\Upsilon$
and by the  large spin splitting between $J/\psi$ and $\eta_c$.
These biases can be avoided by determining $m_Q$ from the masses of heavy mesons and heavy baryons.
An alternative prescription for $m_Q$ is the difference between the sum of the spin centroids of the masses 
for the  doublets of ground-state  mesons with flavors $Q \bar u$ and $Q \bar d$
and the mass of the ground-state baryon with  flavor  $Q[u\, d]$. 
The resulting masses for the charm and bottom quarks  are 
%%%%%%%%
\begin{subequations}
\begin{eqnarray}
m_c \equiv \frac{M_{D^0} + 3 M_{D^{*0}}}{4} + \frac{M_{D^+} + 3 M_{D^{*+}}}{4} - M_{\Lambda_{c[u\, d]}} 
= 1.660~\mathrm{GeV},
\label{mc}
 \\
m_b \equiv \frac{M_{B^0} + 3 M_{B^{*0}}}{4} + \frac{M_{B^+} + 3 M_{B^{*+}}}{4}  - M_{\Lambda_{b[u\, d]}} 
= 5.007~\mathrm{GeV}.
\label{mb-old}
 \end{eqnarray}
\label{mQ}%
\end{subequations}
%%%%%%%%
The masses of $B^{*0}$ and $B^{*+}$ have not been measured separately, 
so they have been set equal to the measured mass $M_{B^*}$.
Using the spin centroids of the heavy meson masses reduces the bias from the
larger hyperfine splittings of the charm mesons.
Contributions to the energies of the light QCD fields that can be  interpreted as 
constituent masses of the  $\bar u$ and $\bar d$ in the heavy mesons 
and the  $u$ and $d$ in the heavy baryon 
are canceled by the subtractions in Eq.~\eqref{mQ}.  
If the heavy hadron masses are expanded to first order in $1/m_Q$ using Eq.~\eqref{H_ell},
our prescription for $m_Q$ implies the constraints
%%%%%%%%
\begin{equation}
\left(\mathcal{E}_{\bar u,1/2^-} + \frac{\mathcal{K}_{\bar u,1/2^-}}{2 m_Q} \right) 
+ \left(\mathcal{E}_{\bar d,1/2^-} + \frac{\mathcal{K}_{\bar d,1/2^-}}{2 m_Q} \right) 
= \mathcal{E}_{[u\, d],0^+} + \frac{\mathcal{K}_{[u\, d],0^+}}{2 m_Q}.
\label{calE-ell}
\end{equation}
%%%%%%%%
Thus the energies $\mathcal{E}_\ell$
are not strictly independent of the heavy quark  mass, but include some resummation of higher orders in $1/m_Q$.

For the lowest-energy states $\ell$ of the light QCD fields,
the coefficients $\mathcal{E}_\ell$, $\mathcal{K}_\ell$, and $\mathcal{S}_\ell$ in the Hamiltonian in Eq.~\eqref{H_ell}
can be deduced from the masses of observed  heavy hadrons containing a single charm or bottom quark.
The Hamiltonian in Eq.~\eqref{H_ell} does not take into account the contributions to the masses 
from electromagnetic interactions between quarks.
The electromagnetic interactions and the $u-d$ mass difference both contribute to the isospin splitting 
between two heavy hadrons that differ only by the replacement of a $u$ quark by a $d$ quark.
The resulting isospin splittings are at most about 5~MeV.
In our analysis, we will not take isospin splittings into account  in the Hamiltonian in Eq.~\eqref{H_ell}.
We do however  take them into account through theoretical errors on hadron masses.
We will try to take into account  
all contributions to hadron masses that  may be larger than the isospin splittings, 
but we will ignore contributions that are much smaller than  10~MeV.

Hyperfine splittings between heavy hadrons come from the $\bm{S} \cdot \bm{j}_\ell$ term in the Hamiltonian in Eq.~\eqref{H_ell}.
The hyperfine splittings between charm mesons are roughly 140~MeV, 
and the hyperfine splittings between bottom mesons are smaller by about a factor of 3.
A correction to the coefficient $\mathcal{S}_{\ell}$ of order $1/m_Q$ could therefore give contributions 
to charm hadron masses that are greater than 10~MeV, and it should therefore be taken into account.  
These corrections can be taken into account by 
replacing $\mathcal{S}_{\ell}$ by a different coefficient $\mathcal{S}_{\ell,Q}$ for the two heavy flavors $Q=c,b$. It is also convenient to absorb the kinetic energy term in the Hamiltonian in Eq.~\eqref{H_ell}
into the coefficient $\mathcal{E}_{\ell}$ by allowing it to be different for the two heavy flavors $Q=c,b$:
%%%%%%%%
\begin{equation}
\mathcal{E}_{\ell,Q} = \mathcal{E}_{\ell}+ \frac{\mathcal{K}_\ell}{2 m_Q} .
\label{E_ell,Q}
\end{equation}
%%%%%%%%
With these two changes, our Hamiltonian for  heavy hadrons with  heavy quark $Q$ 
and  light QCD fields in the state $\ell$ reduces to
%%%%%%%%
\begin{equation}
H^Q_\ell = m_Q + \mathcal{E}_{\ell,Q} 
+ \frac{\mathcal{S}_{\ell,Q}}{2 m_Q} \bm{S} \cdot \bm{j}_\ell.
\label{H_ell,Q}
\end{equation}
%%%%%%%%
An analysis based on this Hamiltonian should be able to predict the masses 
of heavy hadrons with an accuracy of about 5~MeV.

If the coefficients in  the Hamiltonian in Eq.~\eqref{H_ell,Q} have been determined for $Q=c$ and $b$,
they can be extrapolated in $m_Q$ to obtain estimates for the coefficients in  the Hamiltonian in Eq.~\eqref{H_ell}.
The asymptotic coefficient of the kinetic energy term is
%%%%%%%%
\begin{equation}
\mathcal{K}_{\ell} =
\frac{2m_bm_c}{m_b-m_c} \big( \mathcal{E}_{\ell,c} -  \mathcal{E}_{\ell,b} \big).
\label{K_ell,inf}
\end{equation}
%%%%%%%%
The terms proportional to $\mathcal{E}_\ell$ on the right side cancel,
and the terms proportional to $\mathcal{K}_\ell$ add up to $\mathcal{K}_\ell$.
The asymptotic spin-splitting coefficient is
%%%%%%%%
\begin{equation}
\mathcal{S}_{\ell}= \frac{m_b}{m_b-m_c} \mathcal{S}_{\ell,b} - \frac{m_c}{m_b-m_c} \mathcal{S}_{\ell,c}.
\label{S_ell,inf}
\end{equation}
%%%%%%%%
The terms proportional to $\mathcal{S}_\ell$ on the right side add up to $\mathcal{S}_\ell$.
The term proportional to $1/m_b$ in $\mathcal{S}_{\ell,b}$
and the term proportional to $1/m_c$ in $\mathcal{S}_{\ell,c}$ cancel.

The difference between the coefficients $\mathcal{E}_{\ell,Q}$ 
in Eq.~\eqref{H_ell,Q} for states $\ell$ whose light flavor differs only by the replacement of $u$ or $d$ by $s$ 
has a contribution from the mass of the strange quark.
We denote that difference by $m_{s,Q}$, with the dependence on the heavy flavor explicit
but the dependence on the other quantum numbers in $\ell$ suppressed. 
For states $\ell$  that only have flavors  lighter than  $s$,
such as $\bar u$, $[ud]$, or $uu$,
we sometimes also suppress the dependence of $\mathcal{E}_{\ell,Q}$ on the other quantum numbers,
denoting it by $\mathcal{E}_{u/d,Q}$.
We can interpret $m_{s,Q}$ as a constituent mass for the strange quark, 
although it also includes contributions from the difference between the kinetic energies of
$s$ and $u$ or $d$.
The mass splittings between strange and nonstrange hadrons are roughly 100~MeV 
for both charm hadrons and bottom hadrons.  The difference between $m_{s,c}$ and $m_{s,b}$
could be larger than 10~MeV, and it should therefore be taken into account.
The dependence of $m_{s,c}$ and $m_{s,b}$ on the suppressed quantum numbers
for the light QCD state $\ell$ may also be significant.

%%%%%%%%%%%%%%%%%%%%%%%%%%%%%%%%%%%%%%%%%%%%%%%
\subsection{Coefficients for Heavy Mesons from Experiment}
\label{sec:HeavyMesonExperiment}

We proceed to apply the   Hamiltonian in Eq.~\eqref{H_ell,Q}
to the ground-state heavy mesons.
We use the masses of heavy mesons  in the  2020
Review of Particle Physics \cite{Tanabashi:2018oca}.
 The masses have been measured for all the charm mesons in the spin doublets 
$(D^0, D^{*0})$, $(D^+, D^{*+})$, and  $(D_s^+, D_s^{*+})$
and for all the bottom mesons  in the spin doublets 
$(B^-, B^{*-})$, $(\bar B^0, \bar B^{*0})$, and $(\bar B_s^0, \bar B_s^{*0})$,
except that the masses of $B^{*-}$ and $\bar B^{*0}$ are not resolved separately.
 We ignore  the effects of  electromagnetism  and the $u-d$ mass difference
in the Hamiltonian in Eq.~\eqref{H_ell,Q}.
We  instead  treat their contributions to the masses as theoretical errors.
 We use isospin splittings to estimate those theoretical errors.
We obtain the theoretical error in the charm-meson masses 
from the measured $D^+ - D^0$ mass difference and
from the difference between the measured $D^{*+}$ and $D^{*0}$ masses.
Their weighted average is 4.8~MeV. 
We take the theoretical error in the bottom-meson masses to be the 
$\bar B^0 - B^-$ mass difference, which is 0.3~MeV. 
The much smaller isospin splittings for bottom mesons 
can be attributed to a near cancellation between the contributions to the mass from the Coulomb energy 
between the quark and antiquark  and from the $d-u$ mass difference.  
For charm mesons, these two contributions have the same sign.
We obtain the total error in each heavy meson mass by adding the experimental error and the theoretical error linearly.

%%%%%%%%%%%%%%%%%%%%%%%%%%%%%%%%%%%%%%%%%%
\begin{table}[t]
\begin{center}
\begin{tabular}{lc|ccc|cc}
%\rule[10pt]{-1mm}{0mm}
$Q$~ & $\ell$ & ~$\mathcal{E}_{u/d,Q}$ [MeV]~  & ~$m_{s,Q}$ [MeV]~ & ~$\mathcal{S}_{Q}$ [GeV$^2$]~ &
~dof~ & ~$\chi^2$/dof~\\
\hline
$c$    &   $\bar q$, $\tfrac12^-$   & ~$313.4 \pm 2.0$~ & ~$102.3 \pm 3.5$~ & ~$0.472 \pm 0.012$~ &
3 & 0.28  \\
$b$    & ~$\bar q$, $\tfrac12^-$~ &   $306.3 \pm 0.2$   & ~~$87.6 \pm 0.4$     &   $0.455 \pm 0.003$   &
2 & 1.23  \\
 \end{tabular}
\end{center}
\caption{
Coefficients in the Hamiltonian  in Eq.~\eqref{H_ell,Q}  for ground-state heavy mesons
determined from measured meson masses. 
The 3 coefficients in the rows  with heavy flavor $c$ and $b$ were obtained by minimizing the $\chi^2$ 
for 6 charm-meson masses and 5 bottom-meson masses, respectively.
}
\label{tab:Coeffs:hmeson-PDG}
\end{table}
%%%%%%%%%%%%%%%%%%%%%%%%%%%%%%%%%%%%%%%%%%

In a ground-state heavy meson, the light QCD fields are in the state $\ell= \bar q, \frac12^-$.
We determine the coefficients $\mathcal{E}_{u/d,c}$, $m_{s,c}$,  and $\mathcal{S}_{c}$ in Eq.~\eqref{H_ell,Q}
by minimizing the  $\chi^2$ for the masses of the 6 ground-state charm mesons. 
We determine the coefficients $\mathcal{E}_{u/d,b}$, $m_{s,b}$,  and $\mathcal{S}_{b}$
by minimizing the  $\chi^2$ for the 5 measured masses of the  ground-state bottom mesons. 
The results are given in Table~\ref{tab:Coeffs:hmeson-PDG}.
The central values of the coefficients are their values at the minimum of $\chi^2$.
The error bar in a coefficient is  how much it must be changed to 
increase $\chi^2$  by 1 if the other coefficients are held fixed.
The energy $\mathcal{E}_{u/d,c}$ is larger than $\mathcal{E}_{u/d,b}$ 
by about 7~MeV. 
The  strange-quark mass  $m_{s,c}$  is larger than $m_{s,b}$ by about 15~MeV.  
The  coefficient $\mathcal{S}_{c}$  is larger than $\mathcal{S}_{b}$  
by  1.4  error bars.  It corresponds to a difference in the hyperfine splittings 
for charm mesons of about 5~MeV. 

The coefficients  in Table~\ref{tab:Coeffs:hmeson-PDG} are those for the Hamiltonian in Eq.~\eqref{H_ell,Q}.
Extrapolations in $1/m_Q$ 
can be used to estimate the coefficients in the Hamiltonian in Eq.~\eqref{H_ell} for asymptotically large $m_Q$.
The coefficient $\mathcal{K}_\ell$ of the kinetic energy term can be estimated 
by inserting $\mathcal{E}_{u/d,c}$ and $\mathcal{E}_{u/d,b}$
into Eq.~\eqref{K_ell,inf}.  The resulting estimate for 
the kinetic energy $\mathcal{K}_\ell/(2m_c)$  in charm mesons 
with the lightest flavors $\bar u$ and $\bar d$  is about 11~MeV.
By replacing  $\mathcal{E}_{\ell,c} -  \mathcal{E}_{\ell,b}$ in Eq.~\eqref{K_ell,inf} by $m_{s,c}-m_{s,b}$,
the additional kinetic  energy in charm mesons 
with the light flavor $\bar s$  is estimated to be about 22~MeV.
These kinetic energies are not small compared to isospin splittings, so it is important to take them into account.
By inserting $\mathcal{S}_{c}$ and $\mathcal{S}_{b}$ into Eq.~\eqref{S_ell,inf}, 
 the asymptotic coefficient $\mathcal{S}_\ell$ from Eq.~\eqref{S_ell,inf}
is estimated to be $(0.447 \pm 0.008)~\mathrm{GeV}^2$. 
The contribution to the 
 hyperfine splittings of charm mesons from the difference between
$\mathcal{S}_{\ell,c}$ and $\mathcal{S}_\ell$ is $(\mathcal{S}_{\ell,c} -\mathcal{S}_\ell)/2m_c$,
which is about 8~MeV.
This is not small compared to the isospin splittings, so it is important to take it into account.

%%%%%%%%%%%%%%%%%%%%%%%%%%%%%%%%%%%%%%%%%%%%%%%
\subsection{Coefficients for Heavy Baryons from Experiment}
\label{sec:HeavyHadronExperiment}

We proceed to apply the   Hamiltonian in Eq.~\eqref{H_ell,Q}
to the ground-state heavy baryons.
We use the masses of heavy baryons in the 2020
Review of Particle Physics \cite{Tanabashi:2018oca}.
The masses for heavy baryons with antisymmetric light flavor
have been measured for the charm baryons $\Lambda_c^+$, $\Xi_c^+$, and $\Xi_c^0$ 
and the corresponding bottom baryons $\Lambda_b^0$, $\Xi_b^0$,
 and $\Xi_b^-$.
The masses for heavy baryons with symmetric light flavor have been measured for both members of all six
charm-baryon doublets:  $(\Sigma_c^{++}, \Sigma_c^{*++})$, $(\Sigma_c^{+}, \Sigma_c^{*+})$,
$(\Sigma_c^{0}, \Sigma_c^{*0})$, $(\Xi_c^{'+}, \Xi_c^{*+})$, $(\Xi_c^{'0}, \Xi_c^{*0})$,
and $(\Omega_c^{0}, \Omega_c^{*0})$.
They have been measured for both members of three of the six bottom-baryon doublets:  
$(\Sigma_b^+, \Sigma_b^{*+})$, $(\Sigma_b^-, \Sigma_b^{*-})$, and $(\Xi_b^{'-}, \Xi_b^{*-})$.
They have also been measured for $\Xi_b^{*0}$ and $\Omega_b^-$.
We ignore  the effects of  electromagnetism  and the $u-d$ mass difference
in the Hamiltonian in Eq.~\eqref{H_ell,Q}.
We instead  treat their contributions to the masses as theoretical errors.
 We use isospin splittings to estimate those theoretical errors.
There are some isospin splittings for heavy baryons that have been measured directly
with a better precision than the differences between the measured masses.
There are 7 pairs of  ground-state charm baryons  that differ by the replacement of a $u$ quark by a $d$ quark. 
We take the theoretical error in the masses for the charm baryons 
to be the weighted average of the absolute values of these 7 isospin splittings, which is 1.7~MeV. 
The isospin splittings for the ground-state bottom baryons 
that are known are those for $\Xi_b^0-\Xi_b^-$ and $\Xi_b^{*0}-\Xi_b^{*-}$,
which differ by the replacement of a $u$ quark by a $d$ quark,
and those for $\Sigma_b^+-\Sigma_b^-$ and $\Sigma_b^{*+}-\Sigma_b^{*-}$,
which differ by the replacement of two $u$ quarks by two $d$ quarks. 
We take the theoretical error in the masses for the bottom baryons  
to be the weighted average of these 4 isospin splittings, which is 4.9~MeV.
We obtain the total error in each heavy baryon mass by adding the theoretical error and the experimental error  linearly.

In a ground-state heavy baryon with  antisymmetric light flavor $[q\, q']$,
 the light QCD fields are in the state $\ell = [q\, q'],0^+$.
We determine the coefficients $\mathcal{E}_{u/d,c}$ and $m_{s,c}$  in Eq.~\eqref{H_ell,Q}
by minimizing the  $\chi^2$ for the masses of the 3 charm baryons 
$\Lambda_c^+$, $\Xi_c^+$, and $\Xi_c^0$.
We determine the coefficients $\mathcal{E}_{u/d,b}$ and $m_{s,b}$  
by minimizing the  $\chi^2$ for the masses of the 3 bottom baryons 
$\Lambda_b^0$, $\Xi_b^0$, and $\Xi_b^-$.
The results are given in Table~\ref{tab:Coeffs:hbaryon-PDG}.
In a ground-state heavy baryon with symmetric light flavor $\{q\, q'\}$,
 the light QCD fields are in the state $\ell = \{q\, q'\},1^+$.
We determine the coefficients $\mathcal{E}_{u/d,c}$, $m_{s,c}$,  and $\mathcal{S}_{c}$
by minimizing the  $\chi^2$ for the masses of the 12 baryons 
in the six charm-baryon doublets.
We determine the coefficients $\mathcal{E}_{u/d,b}$, $m_{s,b}$,  and $\mathcal{S}_{b}$
by minimizing the  $\chi^2$ for the masses of the 8  baryons in the six
bottom-baryon doublets whose masses have been measured. 
The results are given in Table~\ref{tab:Coeffs:hbaryon-PDG}.

%%%%%%%%%%%%%%%%%%%%%%%%%%%%%%%%%%%%%%%%%%
\begin{table}[t]
\begin{center}
\begin{tabular}{lc|ccc|cc}
%\rule[10pt]{-1mm}{0mm}
$Q$~ & ~$\ell$~ & ~$\mathcal{E}_{u/d,Q}$ [MeV]~ & ~$m_{s,Q}$ [MeV]~ & ~$\mathcal{S}_{Q}$ [GeV$^2$]~  &
~dof~ & ~$\chi^2$/dof~\\
\hline
$c$    &   ~$[q\, q']$, $0^+$    &    $626.5 \pm 1.1$  &   $182.9 \pm 1.4$  &                                      &
1 & 1.12  \\
$b$    &   ~$[q\, q']$, $0^+$    &    $612.4 \pm 3.0$  &    $174.8 \pm 3.8$ &                                      &
1 &  0.45   \\
\hline  
$c$    &    $\{q\, q'\}$, $1^+$   &    $837.7 \pm 0.6$  &   $124.2 \pm 0.8$  &  ~$0.147 \pm 0.003$~ &
9 & 0.92  \\
$b$    & ~$\{q\, q'\}$, $1^+$~  &   $819.8 \pm 1.9$   &   $118.8 \pm 2.2$   &   $0.137 \pm 0.024$    &
5 & 0.44   \\
\hline
\end{tabular}
\end{center}
\caption{
Coefficients in the Hamiltonian  in Eq.~\eqref{H_ell,Q} for ground-state heavy baryons
 determined from measured baryon masses. 
The 2 coefficients in the rows with flavor $c\, [q\, q']$ and $b\, [q\, q']$
were obtained by minimizing the  $\chi^2$ 
for 3 charm-baryon masses and 3 bottom-baryon masses, respectively.
The 3 coefficients in the rows with flavor $c\, \{q\, q'\}$ and $b\, \{q\, q'\}$
were obtained by minimizing the  $\chi^2$ 
for 12 charm-baryon masses and 8 bottom-baryon masses, respectively.
}
\label{tab:Coeffs:hbaryon-PDG}
\end{table}
%%%%%%%%%%%%%%%%%%%%%%%%%%%%%%%%%%%%%%%%%%

The energies $\mathcal{E}_{u/d,Q}$  in Table~\ref{tab:Coeffs:hbaryon-PDG} 
for heavy baryons with antisymmetric light flavor  and symmetric light flavor 
are larger than those for heavy mesons
in Table~\ref{tab:Coeffs:hmeson-PDG} by about 310~MeV and 520~MeV, respectively.
The strange-quark masses $m_{s,Q}$ in Table~\ref{tab:Coeffs:hbaryon-PDG}
for heavy baryons with antisymmetric light flavor  and symmetric light flavor  are
larger than those for heavy  mesons in Table~\ref{tab:Coeffs:hmeson-PDG}
by about 85~MeV and 25~MeV, respectively.
The spin-splitting coefficients $\mathcal{S}_Q$ for heavy baryons with symmetric light flavor
in Table~\ref{tab:Coeffs:hbaryon-PDG} are smaller than those for heavy mesons in Table~\ref{tab:Coeffs:hmeson-PDG} by about a factor of 1/3.

Some of the differences between the coefficients for $Q=c$ and $b$ in Table~\ref{tab:Coeffs:hbaryon-PDG} 
correspond to energy differences that are larger than isospin splittings.
The  energies $\mathcal{E}_{u/d,c}$ are larger than $\mathcal{E}_{u/d,b}$ 
by about 15~MeV.
The strange-quark masses $m_{s,c}$ are larger than $m_{s,b}$ by about 7~MeV.
The  spin-splitting coefficients $\mathcal{S}_{c}$  and $\mathcal{S}_{b}$
are equal to within the errors.

%%%%%%%%%%%%%%%%%%%%%%%%%%%%%%%%%%%%%%%%%%%%%%%
\subsection{Coefficients for Heavy Baryons from Lattice QCD}
\label{sec:HeavyBaryonLattice}

The coefficients in the Hamiltonians  for heavy hadrons in Eq.~\eqref{H_ell,Q}
can also be determined from the masses of hadrons calculated using lattice QCD.
There have been several calculations of the masses of  baryons containing heavy quarks 
using lattice QCD with all  the major sources of systematic uncertainties under control.
The systematic errors include those from the extrapolations to zero lattice spacing, 
to infinite volume, and to the physical light quark masses.
The $b$ quark is always treated using lattice NRQCD,
which introduces additional systematic errors.
Briceno {\it et al.}\ \cite{Briceno:2012wt}, Alexandrou {\it et al.}\ \cite{Alexandrou:2014sha}, 
and Brown {\it et al.}\ \cite{Brown:2014ena} have calculated the masses of ground-state baryons 
containing one, two or three $c$ quarks with dynamical light quarks and with the systematic errors quantified.  
There are also calculations of these masses with only a single lattice spacing 
\cite{Liu:2009jc, Namekawa:2013vu, Bali:2015lka, Alexandrou:2017xwd,  Bahtiyar:2020uuj}. 
Brown {\it et al.}\ have calculated the masses of the ground-state baryons containing one, two, or three  $b$ quarks, 
including those containing $bc$, $bcc$, and $bbc$,
with dynamical light quarks and with the systematic errors quantified \cite{Brown:2014ena}.
There is a previous calculation of the masses of ground-state baryons containing one or two  $b$ quarks
at a single lattice spacing \cite{Lewis:2008fu}.
There is also a recent calculation of the masses of ground-state baryons containing a single  $b$ quark,
including those containing $bc$ and $bcc$, with only statistical errors \cite{Mohanta:2019mxo}.
In all these lattice QCD calculations, the effects of  electromagnetism and the $u-d$ mass difference were ignored.

We proceed to apply the  Hamiltonian in Eq.~\eqref{H_ell,Q}
to the ground-state heavy baryons using masses calculated using lattice QCD.
We include only those calculations in which  all the important systematic errors have been quantified.
For charm baryons, we use the masses calculated in Refs.~\cite{Briceno:2012wt,Alexandrou:2014sha,Brown:2014ena}.
The charm baryons with antisymmetric light flavor are the two spin singlets $\Lambda_c^+$ and $\Xi_c$.
The charm baryons with symmetric light flavor are
the three spin doublets  $(\Sigma_c, \Sigma_c^*)$, $(\Xi_c', \Xi_c^*)$, and $(\Omega_c^{0}, \Omega_c^{*0})$. 
We have suppressed the superscript with the electric charge on members of isospin doublets and isospin triplets.
For bottom baryons, we use the masses calculated by Brown {\it et al.}\ \cite{Brown:2014ena}.
Some mass differences, including all the hyperfine splittings for the 3 charm-baryon doublets 
and the 3 bottom-baryon doublets, were also calculated in Ref.~\cite{Brown:2014ena}.
This is particularly important  for bottom baryons, because the errors in the mass differences are smaller 
than the errors obtained by subtracting the masses by about a factor of 10.
For some charm baryons, the errors in the mass differences are smaller by about a factor of  4.
We obtain a single error for each mass or mass difference calculated using lattice QCD 
by adding the statistical and systematic errors linearly.

%%%%%%%%%%%%%%%%%%%%%%%%%%%%%%%%%%%%%%%%%%
\begin{table}[t]
\begin{center}
\begin{tabular}{cc|rrr|ccl}
%\rule[10pt]{-1mm}{0mm}
~$Q$~ & ~$\ell$~ & ~$\mathcal{E}_{u/d,Q}$ [MeV]~ & ~$m_{s,Q}$ [MeV]~ & ~$\mathcal{S}_Q$ [GeV$^2$]~~~  &
~dof~ & ~$\chi^2$/dof~ & \\
\hline
$c$ &   $[q\, q']$, $0^+$     & $623.7 \pm 17.7$~  & $172.3 \pm 26.5 $~  &                                  & 
4 & 0.12 &   \\
$b$ &   $[q\, q']$, $0^+$     & $618.8 \pm 50.7$~  & $145.0\pm 65.0$~   &                                  & 
0 &        & \\
\hline
$c$ &   $\{q\, q'\}$, $1^+$   & $860.2 \pm 12.8$~ & $102.3 \pm 10.3$~  & ~$0.167 \pm 0.016$~ & 
16 & 0.05 &   \\
$b$ & ~$\{q\, q'\}$, $1^+$~ & $854.2 \pm 29.7$~  & $103.6 \pm 21.5$~      & $0.173 \pm 0.043$~  & 
4 & 0.03 &   \\
 \end{tabular}
\end{center}
\caption{
Coefficients in the Hamiltonian in Eq.~\eqref{H_ell,Q}  for ground-state heavy baryons
 determined from  baryon masses calculated using lattice QCD.
The coefficients $\mathcal{S}_Q$  
were determined from hyperfine splittings of  heavy-baryon doublets.
The coefficients  $\mathcal{E}_{u/d,Q}$  and $m_{s,Q}$ 
 were obtained by minimizing the $\chi^2$ 
of heavy-baryon masses. 
}
\label{tab:Coeffs:hbaryon-lattice}
\end{table}
%%%%%%%%%%%%%%%%%%%%%%%%%%%%%%%%%%%%%%%%%%

The only term in the Hamiltonian in Eq.~\eqref{H_ell,Q}  that contributes to the hyperfine splittings within a heavy-baryon 
spin multiplet is the spin-dependent term with coefficient $\mathcal{S}_{\ell,Q}$.
We determine $\mathcal{S}_{\ell,Q}$ for $\ell =\{q\, q'\}$, $1^+$ by using the hyperfine splittings 
in heavy-baryon doublets calculated  in Ref.~\cite{Brown:2014ena}.
The values of  $\mathcal{S}_c$ and $\mathcal{S}_b$ are determined 
by minimizing the $\chi^2$ for the hyperfine splittings in the 3 charm-baryon doublets  
and the 3 bottom-baryon doublets, respectively.
The results are given in Table~\ref{tab:Coeffs:hbaryon-lattice}.

We determine the coefficients $\mathcal{E}_{u/d,Q}$ and $m_{s,Q}$   in the Hamiltonian in Eq.~\eqref{H_ell,Q} 
for heavy baryons by minimizing the  $\chi^2$ for the masses of heavy baryons calculated using lattice QCD.
For charm baryons with antisymmetric light flavor,
we determine $\mathcal{E}_{u/d,c}$ and $m_{s,c}$ by minimizing the  $\chi^2$ 
for the 6 values of the masses for $\Lambda_c^+$ and $\Xi_c$ calculated in
Refs.~\cite{Brown:2014ena, Briceno:2012wt,Alexandrou:2014sha}. 
For bottom baryons with antisymmetric light flavor,
we determine $\mathcal{E}_{u/d,b}$ and $m_{s,b}$
by minimizing the $\chi^2$ for the masses of $\Lambda_b^0$ and $\Xi_b$
calculated in Ref.~\cite{Brown:2014ena}.
The results are given in Table~\ref{tab:Coeffs:hbaryon-lattice}.
For heavy baryons with symmetric light flavor,
we replace the coefficient $\mathcal{S}_{\ell,Q}$ in Eq.~\eqref{H_ell,Q}
by the central value of $\mathcal{S}_{c}$ or $\mathcal{S}_{b}$
in Table~\ref{tab:Coeffs:hbaryon-lattice}.  We ignore the errors in $\mathcal{S}_{c}$ or $\mathcal{S}_{b}$,
because the errors in the hyperfine splittings used to determined these coefficients 
are small compared to the errors in the masses.
For charm baryons with symmetric light flavor,
we determine  $\mathcal{E}_{u/d,c}$ and $m_{s,c}$
by minimizing the $\chi^2$ for the 18 values of the 6 masses for charm baryons in the spin doublets  calculated in
Refs.~\cite{Briceno:2012wt,Alexandrou:2014sha,Brown:2014ena}.
For bottom baryons with symmetric light flavor,
we determine $\mathcal{E}_{u/d,b}$ and $m_{s,b}$
by minimizing the $\chi^2$ for the masses 
of the 6 bottom baryons in the spin doublets calculated in Ref.~\cite{Brown:2014ena}.
The results are given in Table~\ref{tab:Coeffs:hbaryon-lattice}.

The coefficients $\mathcal{E}_{u/d,Q}$, $m_{s,Q}$, and $\mathcal{S}_Q$ in Table~\ref{tab:Coeffs:hbaryon-lattice} 
are the same for $Q=c$ and $b$ to within the errors. 
The errors for $c$ are  smaller than those for $b$, because the $\chi^2$ is calculated using 
many more charm baryon masses.    
The coefficients 
for  $c\, [q\, q^\prime]$, $b\, [q\, q']$, and $b\, \{q\, q'\}$ in Table~\ref{tab:Coeffs:hbaryon-lattice}
are consistent within their errors with those in Table~\ref{tab:Coeffs:hbaryon-PDG}
that were determined from measured baryon masses.
For $c\, \{q\, q^\prime\}$, the differences between the  coefficients in Table~\ref{tab:Coeffs:hbaryon-lattice} and Table~\ref{tab:Coeffs:hbaryon-PDG}  are more significant.
The coefficient $\mathcal{E}_{u/d,c}$ in Table~\ref{tab:Coeffs:hbaryon-lattice} is larger by  1.8 error bars. 
The coefficient $m_{s,c}$ in Table~\ref{tab:Coeffs:hbaryon-lattice} is smaller by 2.1 error bars. 
The coefficient  $\mathcal{S}_c$ in Table~\ref{tab:Coeffs:hbaryon-lattice}  
is larger by 1.2 error bars.

%\newpage

%%%%%%%%%%%%%%%%%%%%%%%%%%%%%%%%%%%%%%%%%%%%%%%
\section{Doubly Heavy Hadrons}
\label{sec:2Heavycore}
%%%%%%%%%%%%%%%%%%%%%%%%%%%%%%%%%%%%%%%%%%%%%%%

In this section, we consider baryons  and mesons that contain two heavy quarks.
An effective field theory for doubly heavy hadrons  \cite{Brambilla:2005yk,Fleming:2005pd} 
can be used to organize their masses into expansions in inverse powers of the heavy quark masses.
We determine the coefficients in the expansions of the masses of doubly heavy  baryons 
using masses calculated using lattice QCD.
We then use the heavy-quark-diquark symmetry to 
determine coefficients in the expansions of the masses for doubly heavy tetraquarks.
Finally we use these expansions to
predict the masses of doubly heavy tetraquarks.
This section  follows the strategy of Eichten and Quigg in Ref.~\cite{Eichten:2017ffp},
but with some corrections and improvements.

%%%%%%%%%%%%%%%%%%%%%%%%%%%%%%%%%%%%%%%%%%%%%%%
\subsection{ Heavy Diquark}
\label{sec:2Heavy-core}

In a doubly heavy hadron, the heavy diquark has the flavor $Q_1 Q_2$ of two heavy quarks.
Since the two heavy quarks are color triplets ($\bm{3}$), 
the color state of the  heavy diquark is a linear combination of anti-triplet $\left(\bm{3^*}\right)$ 
and sextet $\left(\bm{6}\right)$ states.
If the heavy quarks are separated by a small distance $r$,
they have a large associated potential energy.
If their color state is  $\bm{3^*}$, the color-Coulomb potential at short distances is $-2\alpha_s/3r$, which is 
attractive.
If their color state is  $\bm{6}$, the color-Coulomb potential at short distances is $+\alpha_s/3r$, which is repulsive.
We assume the attractive color-Coulomb potential binds the two heavy quarks 
into a compact diquark in the anti-triplet color state.
We  ignore the color-sextet component of the  heavy diquark.

Given that the color state of the two heavy quarks is $\bm{3^*}$, they are antisymmetric in color.
If the two heavy quarks have the same flavor $Q$, the combination of their spatial and spin states must be symmetric.
The total spin  $\bm{S} = \bm{S}_1 + \bm{S}_2$ of the diquark is the sum of the spins of the two heavy quarks.
The spin state can be antisymmetric with spin quantum number $S=0$ or symmetric  with  $S=1$.
If the two heavy quarks   have orbital-angular-momentum quantum number $L$,
their spatial state is symmetric if $L$ is even and antisymmetric  if $L$ is odd.
Thus $L+S$ must be odd.  
We will consider only the $L=0$ state of the heavy diquark, which presumably has the lowest energy.
In this case, we must have  $S=1$.

If the two heavy quarks have distinct flavors $Q_1$ and $Q_2$, 
the combination of their flavor, spatial, and spin states must be symmetric.
In the flavor-symmetric state $\{ Q_1,Q_2 \}$, $L+S$ must be odd.  
In the flavor-antisymmetric state $[ Q_1,Q_2 ]$, $L+S$ must be even.  
We will consider only the $L=0$ state of the heavy diquark, which presumably has the lowest energy.
In this case,  
the flavor-antisymmetric state $[ Q_1,Q_2 ]$ must have $S=0$ and
the flavor-symmetric state $\{ Q_1,Q_2 \}$ must have  $S=1$.
The  flavor-antisymmetric state $[ Q_1,Q_2 ]$ is expected to be lower in energy
than the flavor-symmetric state $\{ Q_1,Q_2 \}$.

%%%%%%%%%%%%%%%%%%%%%%%%%%%%%%%%%%%%%%%%%%%%%%%
\subsection{Doubly Heavy  Baryon}
\label{sec:2HeavyBaryon}

In the presence of a diquark consisting of two heavy quarks $Q_1Q_2$, 
the light QCD fields can have finite energy only if 
they  can combine with the two heavy quarks
to form a color singlet. This constrains the flavor of the light QCD fields.  
If the color state of the heavy diquark is $\bm{3}^*$,
the simplest possibility is the flavor of a light quark $q$.
The resulting hadron is  a doubly heavy baryon with flavor $Q_1Q_2  q$.
The  lowest-energy states of the light QCD fields are 
discrete states $\ell$ that are the charge conjugates of  
the discrete states $\bar \ell$ in a heavy meson with flavor $Q\bar q$.
 The states $\ell$ can be specified by the light flavor $ q = u, d, s$,  
the angular-momentum/parity quantum numbers $j^P$, 
where $j$ is a half-integer, and a principal quantum number.  
The light QCD states with the lowest principal quantum number 
in a doubly heavy baryon can be labeled by $\ell = q, j^P$.

An effective field theory called potential NRQCD (pNRQCD) for doubly heavy baryons has been developed by 
Brambilla, Vairo, and Rosch (BVR) \cite{Brambilla:2005yk}.
This effective field theory is analogous to pNRQCD for heavy quarkonium \cite{Brambilla:1999xf}.
BVR considered the general case where the two heavy quarks could have different flavors $Q_1$ and $Q_2$ 
and  different  masses $m_{Q_1}$ and $m_{Q_2}$. 
The effective field theory is formulated in terms of a triplet field $T$ 
for the heavy diquark in a $\bm{3^*}$ color state and a  sextet  field $\Sigma$ 
for the heavy diquark in a {\bf 6} color state together with the usual QCD fields for gluons and light quarks.
The Lagrangian can be expanded in powers of the inverse heavy quark mass  $1/m_Q$
and the radius $r$ of the multipole expansion.
In the case of distinct heavy quarks $Q_1$ and $Q_2$, the expansion in powers of the inverse heavy quark mass 
is actually an expansion in powers of $1/m_{Q_1}$, $1/m_{Q_2}$, and $1/(m_{Q_1}+m_{Q_2})$, 
which we will represent collectively by $1/m_Q$.
BVR wrote down the terms in the Lagrangian explicitly at leading order, at first order in  $1/m_Q$, 
and at first order in $r$. 
They determined the coefficients in the effective Lagrangian by matching Green functions with those of 
nonrelativistic QCD (NRQCD). 
There are four terms in the Lagrangian at first order in $1/m_Q$ 
that depend on the  triplet  field $T$ but not on the  sextet field $\Sigma$:
\begin{itemize}
\item
a kinetic term
$T^\dagger \bm{D}^2 T/\big(2(m_{Q_1} + m_{Q_2}) \big)$ for the motion of the heavy quark pair,
\item
a kinetic term  $T^\dagger \bm{\nabla}^2 T/(2\mu_{Q_1Q_2})$ for the relative motion of the two heavy quarks, 
where $\mu_{Q_1Q_2} = m_{Q_1} m_{Q_2}/(m_{Q_1} + m_{Q_2})$  is their reduced mass,
\item
a spin  term $T^\dagger  ( \bm{S}_1/m_{Q_1} +  \bm{S}_2/m_{Q_2}) \cdot \bm{B} \,T$
that depends on the spins $\bm{S}_1$ and $\bm{S}_2$ of the two heavy quarks
and on the chromomagnetic field $\bm{B}$,
\item
an orbital-angular-momentum  term $T^\dagger   \bm{L} \cdot \bm{B}\, T$
that depends on the relative orbital angular momentum $\bm{L}$ of the  two heavy quarks
and has a coefficient  proportional to $(m_{Q_1}^2+ m_{Q_2}^2)/[m_{Q_1}  m_{Q_2} (m_{Q_1} + m_{Q_2})]$.
\end{itemize}

The effective field theory pNRQCD for doubly heavy baryons was also considered by 
Fleming and Mehen (FM) \cite{Fleming:2005pd}.
FM considered only the case of identical heavy flavors $QQ$.
They derived terms in the Lagrangian for the triplet field $T$ by starting from NRQCD
and making a sequence of field transformations.
FM determined explicitly the $T^\dagger  ( \bm{S}_1 +  \bm{S}_2) \cdot \bm{B} \,T/m_Q$ term
in the Lagrangian.  Their result for its coefficient agrees with that of BVR,
but it differs by a factor of 2 from some previous results.

The effective field theory pNRQCD for doubly heavy baryons developed by 
BVR \cite{Brambilla:2005yk} can be used to organize the masses of doubly heavy baryons into
expansions in powers of inverse heavy quark masses.
We assume the terms in the  pNRQCD Lagrangian involving the sextet field $\Sigma$
can be ignored.
From the form of the pNRQCD Lagrangian for the triplet field $T$ and the light QCD fields, 
which includes the four terms itemized above,
we can infer the form of the Hamiltonian $H^{Q_1Q_2}_\ell$
for doubly heavy baryons with light QCD fields in the state $\ell$.
Through first order  in $1/m_Q$,  
the only operators in the Hamiltonian are the heavy-quark spins $\bm{S}_1$ and $\bm{S}_2$,
the relative orbital angular momentum $\bm{L}$ of the two heavy quarks, and the total
angular momentum $\bm{j}_\ell$ of the light QCD fields. The quantum number  for $\bm{j}_\ell$ is the half-integer $j$.
Since we only consider the $L=0$ state of the heavy diquark,  
we can drop the term in the Hamiltonian involving  $\bm{L}$. 
The resulting Hamiltonian  for doubly heavy baryons to first order in  $1/m_Q$  is 
%%%%%%%%
\begin{equation}
H^{Q_1Q_2}_\ell = (m_{Q_1} + m_{Q_2})  + \mathcal{E}_{Q_1Q_2} 
+  \mathcal{E}_\ell + \frac{\mathcal{K}_\ell}{2 (m_{Q_1}+m_{Q_2})} 
+\frac12  \left( \frac{\mathcal{S}_\ell}{2m_{Q_1}} \bm{S}_1 
+ \frac{\mathcal{S}_\ell}{2m_{Q_2}} \bm{S}_2 \right) \cdot \bm{j}_\ell .
\label{HQ1Q2_ell}
\end{equation}
%%%%%%%%
By heavy-quark--diquark symmetry,
the coefficients $\mathcal{E}_\ell$, $\mathcal{K}_\ell$, and $\mathcal{S}_\ell$ in Eq.~\eqref{HQ1Q2_ell}
are identical to those in the Hamiltonian  $H^Q_\ell$  in Eq.~\eqref{H_ell} for heavy mesons. 
The Hamiltonian in Eq.~\eqref{HQ1Q2_ell} depends on the heavy flavor $Q_1Q_2$ 
through the masses $m_{Q_1}$ and  $m_{Q_2}$ and the  diquark energy $\mathcal{E}_{Q_1Q_2}$. 
The kinetic term for the relative motion of the two heavy quarks has been absorbed into $\mathcal{E}_{Q_1Q_2}$.
The prefactor of 1/2 in  the spin-dependent term in Eq.~\eqref{HQ1Q2_ell} takes into account 
that the two heavy quarks  are in an anti-triplet $\left(\bm{3^*}\right)$ color state.
 This factor was first derived using effective field theory in Refs.~\cite{Brambilla:2005yk,Fleming:2005pd}.
The total angular momentum of the doubly heavy 
baryon is $\bm{J} = \bm{S}+ \bm{j}_\ell$, where $\bm{S}= \bm{S}_1 + \bm{S}_2$.
The quantum number for  $\bm{J}$ is the spin $J$ of the doubly heavy baryon.
If the quantum number $S$ for the total spin of the heavy diquark is 0,
the doubly heavy baryon is a spin singlet with spin $J=j$.
If $S$ is 1, the doubly heavy baryons form a spin doublet with spins $J=\tfrac12, \tfrac32$ if $j=\tfrac12$,
and they form a spin triplet with spins $J=j \!-\! 1,j,j \!+\! 1$  if $j \ge \tfrac32$.

We proceed to  simplify the spin-dependent term in the Hamiltonian  in Eq.~\eqref{HQ1Q2_ell}.
If the spins of the heavy quark are in a state with definite total spin quantum number $S$,
the Wigner-Eckart theorem can be used to reduce the sum 
of the two interaction terms to a single term proportional to $\bm{S} \cdot \bm{j}_\ell$.
The result from the Wigner-Eckart theorem is most conveniently expressed in terms of expectation values
of operators in a state of nonzero quantum number $S$ for the total spin of the heavy quarks:
%%%%%%%%
\begin{eqnarray}
\frac12 \left\langle \left( \frac{\mathcal{S}_\ell}{2m_{Q_1}} \bm{S}_1 
+ \frac{\mathcal{S}_\ell}{2m_{Q_2}} \bm{S}_2 \right) \cdot \bm{j}_\ell \right\rangle
&=& \frac{\mathcal{S}_\ell}{2\bm{S}^2}
\left( \frac{\bm{S}_1 \cdot \bm{S}}{2m_{Q_1}} + \frac{\bm{S}_2\cdot \bm{S}}{2m_{Q_2}} \right)
\big\langle \bm{S}  \cdot \bm{j}_\ell \big\rangle
\nonumber\\
&=& \frac{\mathcal{S}_\ell}{8\mu_{Q_1Q_2}}
\big\langle \bm{S}  \cdot \bm{j}_\ell \big\rangle.
\label{Wigner-Eckart}
\end{eqnarray}
%%%%%%%%
In the last step, we used $\bm{S}_i \cdot \bm{S} = \tfrac12 \bm{S} ^2$.
In Ref.~\cite{Eichten:2017ffp}, 
Eichten and Quigg assumed the spin-dependent term could be obtained from that of a heavy hadron 
in Eq.~\eqref{H_ell} by replacing
the heavy quark mass $m_Q$ in the denominator by the total mass $m_{Q_1}+m_{Q_2}$ of the diquark.
They did not take into account the prefactor of 1/2 in  the spin-dependent term in Eq.~\eqref{HQ1Q2_ell},
which comes from the two heavy quarks being in the $\bm{3^*}$ color state.
The denominator of their spin-dependent term was therefore $2(m_{Q_1}+m_{Q_2})$ 
instead of $8\mu_{Q_1Q_2}$ as in Eq.~\eqref{Wigner-Eckart}.
There is a fortuitous agreement between these two denominators in the case of equal-mass heavy quarks.
In the case of the unequal quark masses for $b$ and $c$ in Eqs.~\eqref{mQ},
the correct  spin-dependent energy is larger by a factor of  1.34.

The spin-dependent term in the Hamiltonian for heavy hadrons is large enough that corrections
to the spin-splitting coefficient $\mathcal{S}_\ell$ of order $1/m_Q$ can give contributions to masses 
that are larger than isospin splittings.  
We can allow for dependence of the spin-splitting  coefficient for a doubly heavy baryon on the heavy flavors $Q_1Q_2$
by replacing  $\mathcal{S}_\ell$ in Eq.~\eqref{Wigner-Eckart} by $\mathcal{S}_{\ell,Q_1Q_2}$.
The dependence on the heavy flavors includes that from matching between pNRQCD and NRQCD 
at higher orders in the $1/m_Q$ expansion.  
The pNRQCD Lagrangian at order $1/m_Q$ includes the terms $T^\dagger \bm{S}_1 \cdot \bm{B} \,T/m_{Q_1}$ 
and $T^\dagger \bm{S}_2 \cdot \bm{B} \,T/m_{Q_2}$ with equal coefficients.
At order $1/m_Q^2$, there are corrections to the coefficients of these two terms proportional to $1/m_{Q_1}$  
and $1/m_{Q_2}$, respectively. 
These corrections are the same as for the corresponding term in HQET for a singly heavy hadron.  
They can therefore be taken into account by replacing $\mathcal{S}_\ell$ in the 
coefficients of the two spin-dependent terms in Eq.~\eqref{HQ1Q2_ell} 
by $\mathcal{S}_{\ell,\bar Q_1}$ and $\mathcal{S}_{\ell,\bar Q_2}$, respectively.
The spin-splitting coefficient  $\mathcal{S}_{\ell, Q_1Q_2}$ in the Hamiltonian 
can then be deduced from the Wigner-Eckardt theorem, as in Eq.~\eqref{Wigner-Eckart}.
The spin-splitting  coefficient for $cc$ and $bb$ baryons are 
the same as for $c$ and $b$ mesons, respectively:
%%%%%%%%
\begin{equation}
\mathcal{S}_{\ell, QQ} =  \mathcal{S}_{\ell, \bar Q} =  \mathcal{S}_{\bar \ell, Q}.
\label{S,QQ}
\end{equation}
%%%%%%%%
The spin-splitting  coefficient for a $bc$ baryon is 
%%%%%%%%
\begin{equation}
\mathcal{S}_{\ell, bc} = 
\frac{\mu_{bc}}{m_b} \, \mathcal{S}_{\ell, \bar b} + \frac{\mu_{bc}}{m_c} \, \mathcal{S}_{\ell, \bar c}.
\label{S,bc}
\end{equation}
%%%%%%%%
Note that this prediction for $\mathcal{S}_{\ell, bc}$ is between $\mathcal{S}_{\ell, \bar b}$ and $\mathcal{S}_{\ell, \bar c}$.

In the Hamiltonian for doubly heavy baryons in Eq.~\eqref{HQ1Q2_ell},
it is  convenient to combine the diquark energy $\mathcal{E}_{Q_1Q_2}$ 
with the energies $\mathcal{E}_{\ell}$ and $\mathcal{K}_\ell/[2( m_{Q_1} + m_{Q_2})]$ of the light QCD fields
into a coefficient $\mathcal{E}_{\ell,Q_1Q_2}$ that depends on the heavy flavors:
%%%%%%%%
\begin{equation}
\mathcal{E}_{\ell,Q_1Q_2} = \mathcal{E}_{Q_1Q_2}+ \mathcal{E}_{\ell}+ \frac{\mathcal{K}_\ell}{2 (m_{Q_1}+m_{Q_2})}  .
\label{E_ell,Q1Q2}
\end{equation}
%%%%%%%%
We  allow for dependence of the spin-splitting coefficients on the heavy flavors $Q_1Q_2$
by replacing  $\mathcal{S}_\ell$ in Eq.~\eqref{Wigner-Eckart} by $\mathcal{S}_{\ell,Q_1Q_2}$.
With these two changes, our Hamiltonian for doubly heavy  baryons in Eq.~\eqref{HQ1Q2_ell}
reduces to the same form as that for singly heavy  hadrons  in Eq.~\eqref{H_ell,Q}:
%%%%%%%%
\begin{equation}
H^{Q_1Q_2}_\ell = (m_{Q_1} + m_{Q_2})  + \mathcal{E}_{\ell,Q_1Q_2} 
+ \frac{\mathcal{S}_{\ell,Q_1Q_2} }{8\mu_{Q_1Q_2}} \bm{S} \cdot \bm{j}_\ell.
\label{H_ell,Q1Q2}
\end{equation}
%%%%%%%% 

The quantum number $S$ for the total spin $\bm{S}$ of the two heavy quarks
depends on the flavor symmetry of the diquark.
If the $L=0$ diquark has the symmetric heavy flavor $\{Q_1,Q_2 \}$, it must be in a symmetric spin state with  $S=1$.
The ground-state for the doubly heavy baryon with a given light flavor $q$ is therefore a  spin doublet
consisting of two states with 
$J^P =\tfrac12^+,\tfrac32^+$. For the identical heavy flavors $QQ$,
the two spin doublets are $(\Xi_{QQ},\Xi_{QQ}^*)$ and $(\Omega_{QQ}, \Omega_{QQ}^*)$. 
For the symmetric heavy flavor  $\{b\, c\}$, the two spin doublets are  $(\Xi'_{bc},\Xi_{bc}^*)$ 
and $(\Omega'_{bc}, \Omega_{bc}^*)$, where the prime on the spin-$\tfrac12$ member of the doublet 
distinguishes it from the corresponding spin-singlet states $\Xi_{bc}$ and $\Omega_{bc}$.
If the $L=0$ diquark has the antisymmetric heavy flavor $[ Q_1,Q_2 ]$, it must be in a antisymmetric spin state with $S=0$.
The ground-state for the doubly heavy baryon with a given light flavor $q$
is therefore a  spin singlet with $J^P=\tfrac12^+$.
For the antisymmetric heavy flavor  $[b\, c]$, the spin-singlet baryons are 
 $\Xi_{bc}$ with light flavor $u$ or $d$ and $\Omega_{bc}$ with light flavor $s$.

Heavy-quark--diquark symmetry gives simple predictions for the hyperfine splittings of $QQ$ baryons.
The predictions are encapsulated in the equality between the spin-splitting coefficients
$\mathcal{S}_{\ell, QQ}$ and $\mathcal{S}_{\bar \ell, Q}$  in Eq.~\eqref{S,QQ}.  
They imply, for example,
%%%%%%%%
\begin{equation}
M_{\Xi_{cc}^{*++}} -  M_{\Xi_{cc}^{++}} =  \frac34 \big( M_{D^{*0}} -  M_{D^0} \big).
\label{M*-M:HQD}
\end{equation}
%%%%%%%%
Our identity between $\mathcal{S}_{\ell, QQ}$ and $\mathcal{S}_{\bar \ell, Q}$  in Eq.~\eqref{S,QQ}
takes into account the $1/m_Q$ corrections in the matching between pNRQCD for doubly heavy baryons and NRQCD.
However there are other contributions to the hyperfine splittings.  
Mehen and Mohapatra have calculated
perturbative and nonperturbative corrections to the 
hyperfine splittings for doubly heavy baryons with identical heavy flavors  \cite{Mehen:2019cxn}. 
The corrections  can be interpreted as contributions to  
the spin-splitting coefficient  
$\mathcal{S}_{\ell, QQ}$ in Eq~\eqref{H_ell,Q1Q2}.
The perturbative correction arises from an effective five-point contact operator coupling four heavy quarks and a gluon. 
It gives a contribution to $\mathcal{S}_{\ell, QQ}$
of order $\alpha_s^2$, where $\alpha_s$ is the strong coupling constant evaluated at the scale $m_Qv$ 
of the relative momentum of the two heavy quarks in the doubly heavy hadron. 
The nonperturbative correction arises from an expansion to next-to-next-to-leading order in the 
inverse heavy quark mass $1/m_Q$.
It gives a contribution to   $\mathcal{S}_{\ell, QQ}$
of order $\Lambda_{\rm QCD}^2/m_Q^2$.
These corrections to the spin-splitting coefficient $\mathcal{S}_\ell$ are taken into account in 
 $\mathcal{S}_{\ell, QQ}$ by allowing it to depend on the 
 heavy flavor. Their dependence on the heavy quark mass differs from
that of the $1/m_Q$ correction from matching between pNRQCD and NRQCD at order $1/m_Q^2$.
The results of Ref.~\cite{Mehen:2019cxn} suggest that the deviations of $\mathcal{S}_{\ell,Q_1Q_2}$
from the asymptotic 
spin-splitting coefficient $\mathcal{S}_{\bar \ell}$ in a heavy meson may have no simple dependence on the heavy quark masses.

%%%%%%%%%%%%%%%%%%%%%%%%%%%%%%%%%%%%%%%%%%%%%%%
\subsection{Doubly Heavy Tetraquark}
\label{sec:2HeavyTetraquark}

The second simplest possibility for the flavor of the light QCD fields in the presence of  a 
heavy diquark in the  color state $\bm{3}^*$ is the flavor $\bar q \bar q'$  of two light antiquarks.
The resulting hadron is  a doubly heavy tetraquark with flavor $Q_1Q_2  \bar q \bar q'$.
The lowest-energy states of the light QCD fields are discrete states $\ell$ that are the charge conjugates of the discrete states $\bar \ell$ in a heavy baryon with flavor $Qqq'$.
The  states $\ell$ can be specified by the light flavor, which can be antisymmetric 
$[\bar q\, \bar q^\prime]$ or symmetric  $\{\bar q \, \bar q^\prime \}$,
the angular-momentum/parity quantum numbers $j^P$, 
where $j$ is an integer, and a  principal quantum number. 
The light QCD states $\ell$ with the lowest principal quantum number 
in a doubly heavy tetraquark can  be labeled by
$[\bar q\, \bar q^\prime], j^P$ or $\{\bar q\, \bar q^\prime\}, j^P$.

The effective field theory pNRQCD for doubly heavy baryons developed by 
Brambilla, Vairo, and Rosch in Ref.~\cite{Brambilla:2005yk} applies equally well to doubly heavy tetraquarks.
The effective field theory can be used to organize the masses 
of doubly heavy tetraquarks into expansions in powers of inverse heavy quark masses.
From the form of the pNRQCD Lagrangian for the triplet field $T$ and the light QCD fields, 
we can infer the form of the Hamiltonian $H^{Q_1Q_2}_\ell$
for doubly heavy tetraquarks with light QCD fields in the state $\ell$.
Through  first order  in $1/m_Q$, 
the only operators in the Hamiltonian are the heavy-quark spins $\bm{S}_1$ and $\bm{S}_2$,
the relative orbital angular momentum $\bm{L}$ of the heavy quarks, 
and the total angular momentum $\bm{j}_\ell$ of the light QCD fields. 
The quantum number for $\bm{j}_\ell$ is the integer $j$.
We consider only the $L=0$ state of the heavy diquark,
so the terms involving $\bm{L}$ can be omitted.
In this case, the Hamiltonian $H^{Q_1Q_2}_\ell$ for doubly heavy tetraquarks  reduces to  Eq.~\eqref{HQ1Q2_ell}.
By heavy-quark--diquark symmetry,
the coefficients $\mathcal{E}_\ell$, $\mathcal{K}_\ell$, and $\mathcal{S}_\ell$
are identical to those in the Hamiltonian  $H^Q_\ell$  in Eq.~\eqref{H_ell}  for heavy baryons.
We can allow for dependence of the coefficients of the spin-dependent terms 
in Eq.~\eqref{HQ1Q2_ell} on the heavy flavor $Q_1Q_2$
by replacing $\mathcal{S}_\ell$ in Eq.~\eqref{Wigner-Eckart} by $\mathcal{S}_{\ell,Q_1Q_2}$.
It is also convenient to combine the diquark energy  $\mathcal{E}_{Q_1Q_2}$
with the energies $\mathcal{E}_{\ell}$ and
$\mathcal{K}_\ell/[2( m_{Q_1} + m_{Q_2})]$ of the light QCD fields
into a coefficient $\mathcal{E}_{\ell,Q_1Q_2}$ that depends on the heavy flavors,
as in Eq.~\eqref{E_ell,Q1Q2}.
The resulting Hamiltonian for doubly heavy tetraquarks  reduces to Eq.~\eqref{H_ell,Q1Q2}.

The quantum number $S$ for the total spin $\bm{S}$ of the two heavy quarks depends on 
the flavor symmetry of the heavy diquark.
If the $L=0$ diquark has the antisymmetric heavy flavor $[ b\,c ]$,  
it must be in a antisymmetric  spin state with $S=0$.
The doubly heavy tetraquarks with light QCD states $\ell =[\bar q\, \bar q'],0^+$
and $\ell=\{\bar q\, \bar q' \},1^+$ are 
spin singlets with $J^P=0^+$ and $J^P=1^+$, respectively.
If the $L=0$ diquark has the symmetric heavy flavor $cc$, $\{b\,c \}$, or $bb$, 
it must be in a symmetric spin state with $S=1$.
The doubly heavy tetraquark with light QCD state $\ell =[\bar q\, \bar q'],0^+$
is a spin singlet with  $J^P=1^+$.
The doubly heavy tetraquarks with light QCD states $\ell=\{\bar q\, \bar q' \},1^+$ are 
a spin triplet with  $J^P=0^+,1^+,2^+$.

Using heavy-quark--diquark symmetry,
the coefficients in the Hamiltonian in Eq.~\eqref{H_ell,Q1Q2} for doubly heavy tetraquarks
can be deduced from the corresponding coefficients for doubly heavy baryons
together with coefficients in the Hamiltonian in Eq.~\eqref{H_ell,Q} for heavy baryons and for heavy mesons.
The spin-splitting coefficient $\mathcal{S}_{\ell,Q_1Q_2}$ for doubly heavy tetraquarks with light QCD fields $\ell$ 
can be determined to first order in $1/m_Q$ from the spin-splitting coefficients 
$\mathcal{S}_{\ell,\bar c}$ and $\mathcal{S}_{\ell,\bar b}$ for heavy antibaryons. 
The spin-splitting coefficients for $cc$, $bb$, and $bc$ tetraquarks are given 
 in   Eqs.~\eqref{S,QQ} and \eqref{S,bc}.
The energy $\mathcal{E}_{\ell,Q_1Q_2}$ for a $Q_1Q_2$
tetraquark with light QCD fields $\ell$ 
can be determined from the energy 
$\mathcal{E}_{\ell^\prime,Q_1Q_2}$ for a $Q_1Q_2$  baryon with light QCD fields $\ell'$ 
together with the  energies $\mathcal{E}_{\ell,\bar c}$ and $\mathcal{E}_{\ell,\bar b}$
for heavy antibaryons and the  energies $\mathcal{E}_{\ell^\prime,\bar c}$ 
and $\mathcal{E}_{\ell^\prime,\bar b}$ for heavy mesons. 
The energies $\mathcal{E}_{\ell,Q_1Q_2}$ for $cc$, $bb$, and $bc$ tetraquarks
 through  first order in $1/m_Q$ are
%%%%%%%%%%%%%%%%%%%%%%%%%%%%%%%%%%%
\begin{subequations}
\begin{eqnarray}
\mathcal{E}_{\ell,cc} &=& \mathcal{E}_{\ell^\prime,cc} +
\frac{m_b -2m_c}{2(m_b-m_c)} \big( \mathcal{E}_{\ell,\bar c} - \mathcal{E}_{\ell',\bar c} \big)
+ \frac{m_b}{2(m_b-m_c)} \big( \mathcal{E}_{\ell,\bar b} - \mathcal{E}_{\ell',\bar b} \big),
\label{E_ell,cc}
\\
\mathcal{E}_{\ell,bb} &=& \mathcal{E}_{\ell^\prime,bb} 
-\frac{m_c}{2(m_b-m_c)} \big( \mathcal{E}_{\ell,\bar c} - \mathcal{E}_{\ell',\bar c} \big)
+ \frac{2m_b-m_c}{2(m_b-m_c)} \big( \mathcal{E}_{\ell,\bar b} - \mathcal{E}_{\ell',\bar b} \big),
\label{E_ell,bb}
\\
\mathcal{E}_{\ell,bc} &=& \mathcal{E}_{\ell^\prime,bc} 
-\frac{m_c^2}{m_b^2-m_c^2}\, \big( \mathcal{E}_{\ell,\bar c} - \mathcal{E}_{\ell',\bar c} \big)
+ \frac{m_b^2}{m_b^2-m_c^2} \, \big( \mathcal{E}_{\ell,\bar b} - \mathcal{E}_{\ell',\bar b} \big).
\label{E_ell,bc}
\end{eqnarray}
\label{E_ell,cc,bb,bc}%
\end{subequations}
%%%%%%%%%%%%%%%%%%%%%%%%%%%%%%%%%%%5
The diquark energy $\mathcal{E}_{Q_1Q_2}$ cancels between $\mathcal{E}_{\ell,Q_1Q_2}$
and $ \mathcal{E}_{\ell^\prime,Q_1Q_2}$. 
The coefficients of $\mathcal{E}_{\ell}$ and $\mathcal{E}_{\ell'}$  on the left sides and right sides match.
The coefficients of $\mathcal{K}_{\ell}$ and $\mathcal{K}_{\ell'}$  on the left sides and right sides also match.
The strange-quark masses $m_{s,\ell,Q_1Q_2}$ for  $Q_1Q_2$ tetraquarks 
are differences between  energies  in Eq.~\eqref{E_ell,Q1Q2}
obtained by replacing a $u$ or $d$ quark by a $s$ quark. 
Since the heavy-diquark energy cancels in the difference, $m_{s,\ell,Q_1Q_2}$
can be determined  from strange-quark masses in heavy anti-baryons 
without the subtractions in Eq.~\eqref{E_ell,cc,bb,bc}.
The strange-quark masses for $cc$, $bb$, and $bc$ tetraquarks
through first order in $1/m_Q$ are
%%%%%%%%%%%%%%%%
\begin{subequations}
\begin{eqnarray}
m_{s,\ell, cc} &=&\frac{m_b-2m_c}{2(m_b-m_c)} \, m_{s,\ell, \bar c }  + \frac{m_b}{2(m_b-m_c)}  \, m_{s,\ell, \bar b},
\\
m_{s,\ell, bb} &=& 
-\frac{m_c}{2(m_b-m_c)} \, m_{s,\ell, \bar c }  +  \frac{2m_b-m_c}{2(m_b-m_c)}  \, m_{s,\ell, \bar b},
\\
m_{s,\ell, bc} &=& 
-\frac{m_c^2}{m_b^2-m_c^2} \, m_{s,\ell, \bar c }  + \frac{m_b^2}{m_b^2-m_c^2}  \, m_{s,\ell, \bar b}.
\label{ms,bc}
\end{eqnarray}
\label{ms,cc,bb,bc}
\end{subequations}
%%%%%%%%

%%%%%%%%%%%%%%%%%%%%%%%%%%%%%%%%%%%%%%%%%%%%%%%
\subsection{Coefficients for Doubly Heavy Baryons from Lattice QCD}
\label{sec:DoublyHeavyBaryonLattice}

The coefficients in the Hamiltonian  in Eq.~\eqref{H_ell,Q1Q2} for doubly heavy  baryons
can be determined from the masses of doubly heavy baryons calculated using lattice QCD.
There have been several calculations of the masses of doubly heavy baryons
using lattice QCD with all  the major sources of systematic uncertainties under control.
The systematic errors include those from the extrapolations to zero lattice spacing, 
to infinite volume, and to the physical light quark masses.
In $bc$ and $bb$ baryons, there are additional systematic errors from using lattice NRQCD for the $b$ quark.
Briceno {\it et al.}\ \cite{Briceno:2012wt}, Alexandrou {\it et al.}\ \cite{Alexandrou:2014sha}, 
and Brown {\it et al.}\ \cite{Brown:2014ena} have calculated the masses of all the ground-state $cc$ baryons
with dynamical light quarks and with the systematic errors quantified.
There are also calculations of these masses with only a single lattice spacing  
\cite{Namekawa:2013vu, Bali:2015lka, Alexandrou:2017xwd, Bahtiyar:2020uuj}. 
Mathur and Padmanath have  calculated  the masses of $\Omega_{cc}$ and $\Omega^*_{cc}$  
with the systematic errors quantified  \cite{Mathur:2018rwu}.
Brown {\it et al.}\ have calculated the masses of all the  ground-state $bb$ and $bc$ baryons with dynamical light quarks and with the systematic errors quantified \cite{Brown:2014ena}. 
There is a previous calculation of the masses of ground-state $bb$ baryons at a single lattice spacing \cite{Lewis:2008fu}.
There is also a recent calculation of the masses of the ground-state $bc$ baryons 
with only statistical errors \cite{Mohanta:2019mxo}. 
In all these lattice QCD calculations, the effects of  electromagnetism and the $u-d$ mass difference were ignored.

We proceed to apply the  Hamiltonian in Eq.~\eqref{H_ell,Q1Q2}
to the ground-state doubly heavy baryons using masses calculated using lattice QCD.
We include only those calculations in which  all the important systematic errors have been quantified.
For $cc$ baryons, we use the masses calculated in 
Refs.~\cite{Briceno:2012wt,Alexandrou:2014sha,Brown:2014ena,Mathur:2018rwu}.
For $bc$ and $bb$ baryons, we use the masses calculated by Brown {\it et al.}\ \cite{Brown:2014ena}.
Some mass differences, including all the hyperfine splittings within the two doublets of $cc$ baryons, 
the two doublets of $bc$ baryons, and the two doublets of $bb$ baryons, 
were also calculated in Ref.~\cite{Brown:2014ena}.
Mathur and Padmanath calculated the hyperfine splitting between $\Omega^*_{cc}$ 
and $\Omega_{cc}$ \cite{Mathur:2018rwu}.
The calculations of the hyperfine splittings is important, because the errors in the mass differences 
are 5 to 10 times smaller than the errors that would be obtained by subtracting the masses.

%%%%%%%%%%%%%%%%%%%%%%%%%%%%%%%%%%%%%%%%%%
\begin{table}[t]
\begin{center}
\begin{tabular}{cc|ccc|ccl}
%\rule[10pt]{-1mm}{0mm}
~$Q_1Q_2$~&~$\ell$~&~$\mathcal{E}_{u/d,Q_1Q_2}$ [MeV]~&~$m_{s,Q_1Q_2}$ [MeV]~&
~$\mathcal{S}_{Q_1Q_2}$ [GeV$^2$]~ &~dof~&~$\chi^2$/dof~& \\
\hline
$c\, c$      & $q$, $\tfrac12^+$ & $319.5 \pm 11.0$ & $124.9 \pm 13.4$ & $0.363 \pm 0.024$ & 12 & 0.29 & \\
$[b\, c]$    & $q$, $\tfrac12^+$ & $275.8 \pm 37.2$ &   $55.0 \pm 47.0$ &                                 & 0 &   &      \\
$\{b\, c\}$ & $q$, $\tfrac12^+$ & $309.3 \pm 27.3$ &    ~~$73.5 \pm 34.3$ & $0.181 \pm 0.046$ & 2 & $8\times 10^{-5}$ &   \\
$b\, b$     & $q$, $\tfrac12^+$ & $152.0 \pm 25.1$ &  $130.0 \pm 33.6$ & $0.472 \pm 0.075$ & 2 & $2\times 10^{-5}$ &   \\
 \end{tabular}
\end{center}
\caption{
Coefficients in the Hamiltonian in Eq.~\eqref{H_ell,Q1Q2} for ground-state doubly heavy baryons
determined from baryon masses calculated using lattice QCD. 
The values of  $\mathcal{S}_{Q_1Q_2}$ in the rows 
 with heavy flavor $c\, c$, $\{b\, c\}$, and $b\, b$
were determined by minimizing the $\chi^2$ for 3, 2, and 2  hyperfine splittings, respectively.
The  values of $\mathcal{E}_{u/d,Q_1Q_2}$ and $m_{s,Q_1Q_2}$
in the rows  with heavy flavor  $c\, c$, $[b\, c]$, $\{b\, c\}$, and $b\, b$  
were obtained by minimizing the $\chi^2$ for 14, 2, 4, and  4 doubly heavy baryon masses, respectively.
}
\label{tab:Coeffs:dhbaryon-lattice}
\end{table}
%%%%%%%%%%%%%%%%%%%%%%%%%%%%%%%%%%%%%%%%%%

The only term in the Hamiltonian for doubly heavy baryons  in Eq.~\eqref{H_ell,Q1Q2}  
that contributes to the hyperfine splittings within a  
spin multiplet is the spin-dependent term with coefficient $\mathcal{S}_{\ell,Q_1Q_2}$.
We determine $\mathcal{S}_{\ell,Q_1Q_2}$ for $\{Q_1,Q_2\}$ by using hyperfine splittings 
in doubly heavy-baryon doublets calculated using lattice QCD.
The value of $\mathcal{S}_{cc}$ is determined by minimizing the $\chi^2$ for the 3 results 
for the hyperfine splittings in the two  doublets 
$(\Xi_{cc},\Xi_{cc}^*)$ and $(\Omega_{cc}, \Omega_{cc}^*)$
calculated in  Refs.~\cite{Brown:2014ena} and \cite{Mathur:2018rwu}.
The value of $\mathcal{S}_{bb}$ is determined by minimizing the $\chi^2$ for the hyperfine splittings 
in the two   doublets 
$(\Xi_{bb},\Xi_{bb}^*)$ and $(\Omega_{bb}, \Omega_{bb}^*)$ calculated in Ref.~\cite{Brown:2014ena}. 
The value of  $\mathcal{S}_{\{ bc \}}$ is determined by minimizing the $\chi^2$ 
for the hyperfine splittings in the  two  doublets 
$(\Xi'_{bc},\Xi_{bc}^*)$ and $(\Omega'_{bc}, \Omega_{bc}^*)$
calculated in Ref.~\cite{Brown:2014ena}. 
The results are given in Table~\ref{tab:Coeffs:dhbaryon-lattice}.

We determine the coefficients $\mathcal{E}_{u/d,Q_1Q_2}$ and $m_{s,Q_1Q_2}$   
in the Hamiltonian for doubly heavy baryons in Eq.~\eqref{H_ell,Q1Q2}  by minimizing the $\chi^2$ for the difference 
between the  predicted masses and the masses of doubly heavy baryons calculated using lattice QCD.
The statistical and systematic errors in each mass calculated using lattice QCD are added linearly.
In the Hamiltonian for heavy baryons with symmetric heavy flavor, 
we replace the coefficient  $\mathcal{S}_{\ell,Q_1Q_2}$ 
by the central value of $\mathcal{S}_{cc}$,  $\mathcal{S}_{\{ bc \}}$,  or $\mathcal{S}_{bb}$
in Table~\ref{tab:Coeffs:dhbaryon-lattice}.  We ignore the errors in $\mathcal{S}_{\ell,Q_1Q_2}$,
because the errors in the hyperfine splittings used to determine these coefficients 
are small compared to the errors in the masses.
In the  Hamiltonian for $cc$ baryons, we determine $\mathcal{E}_{u/d,cc}$ and $m_{s,cc}$
by minimizing the $\chi^2$ for the 14 values of the four $cc$ baryon masses calculated in
Refs.~\cite{Briceno:2012wt,Alexandrou:2014sha,Brown:2014ena,Mathur:2018rwu}.
In the Hamiltonians for $[bc]$, $\{b,c\}$ and $bb$  baryons,
we determine the coefficients $\mathcal{E}_{u/d,Q_1Q_2}$ and $m_{s,Q_1Q_2}$ by minimizing the $\chi^2$ 
for the 2, 4, and 4 masses calculated in Ref.~\cite{Brown:2014ena}, respectively.
The results are all given in Table~\ref{tab:Coeffs:dhbaryon-lattice}.

Our results for the energies of light QCD fields in the state $\ell =q,\tfrac12^+$ in 
Table~\ref{tab:Coeffs:dhbaryon-lattice} for doubly heavy baryons 
can be compared to those  of light QCD fields in the charge-conjugate state 
$\bar \ell =\bar q,\tfrac12^-$  in Table~\ref{tab:Coeffs:hmeson-PDG} for heavy mesons.
The  energies $\mathcal{E}_{u/d,Q_1Q_2}$ and $\mathcal{E}_{u/d,Q}$ 
cannot be compared, because $\mathcal{E}_{u/d,Q_1Q_2}$  includes the diquark energy
$\mathcal{E}_{Q_1Q_2}$. 
The values of the strange-quark mass in Table~\ref{tab:Coeffs:dhbaryon-lattice}
are consistent with $m_s$  in Table~\ref{tab:Coeffs:hmeson-PDG} averaged over the heavy flavors $Q=b,c$,
although the errors on $m_s$ for $\{b\, c\}$, $b\, b$, and $[b\, c]$ are very large.
The  spin-splitting coefficient  $\mathcal{S}_{bb}$ in Table~\ref{tab:Coeffs:dhbaryon-lattice}  
is consistent within the errors with  $\mathcal{S}_b$ in Table~\ref{tab:Coeffs:hmeson-PDG},
in accord with Eq.~\eqref{S,QQ}.
The value of $\mathcal{S}_{cc}$ in Table~\ref{tab:Coeffs:dhbaryon-lattice} is smaller than that of $\mathcal{S}_c$ in 
Table~\ref{tab:Coeffs:hmeson-PDG} by about 4 error bars.
This raises  a question about the accuracy of the compact heavy-diquark approximation for $cc$ baryons.
The value of $\mathcal{S}_{\{b\, c\}}$ in Table~\ref{tab:Coeffs:dhbaryon-lattice} 
is only about one half of  $\mathcal{S}_{cc}$,
which is incompatible with the prediction in Eq.~\eqref{S,bc} that $\mathcal{S}_{\{b\, c\}}$
is between $\mathcal{S}_{cc}$ and $\mathcal{S}_{bb}$.
The value of $\mathcal{S}_{\{b\, c\}}$ is smaller than the prediction in Eq.~\eqref{S,bc} by 6.1 error bars.
It  is however consistent with $\mathcal{S}_b$ for a heavy baryon in Table~\ref{tab:Coeffs:hbaryon-PDG}.
This suggests that it may be  more appropriate to regard the charm quark in a $bc$ baryon as 
a light quark instead of as a constituent of a $bc$ diquark.

The  first definitive discovery of a double-charm baryon  was  that of the $\Xi_{cc}^{++}$ 
by the LHCb collaboration in 2017 \cite{Aaij:2017ueg}.
Its  mass is measured to be  $(3621.2 \pm 0.7)$~MeV \cite{Aaij:2017ueg,Aaij:2018gfl}.
(We ignore the previous observations of double-charm baryons 
with smaller masses by the SELEX collaboration \cite{Mattson:2002vu,Engelfried:2005kd}.)
We can combine the 3 results on the  $\Xi_{cc}^{++}$ mass from lattice QCD in
Refs.~\cite{Briceno:2012wt,Alexandrou:2014sha,Brown:2014ena} to give a postdiction for the mass.
If the statistical and systematic errors in each mass are added  linearly, 
the predicted mass  from averaging the 3 results is $(3585 \pm 25)$~MeV.
 This is 1.4 error bars below the mass measured by LHCb.
Alternatively, the  mass of the $\Xi_{cc}^{++}$ can be predicted using the Hamiltonian in Eq.~\eqref{H_ell,Q1Q2}.
The expression for the mass is
%%%%%%%%
\begin{equation}
M_{\Xi_{cc}^{++}} = 2 m_c  + \mathcal{E}_{u/d,cc}  - \mathcal{S}_{cc} /(4m_c).
\label{M_Xicc}
\end{equation}
%%%%%%%%
Upon inserting the values of $m_c$ from Eq.~\eqref{mc}
and $\mathcal{E}_{u/d,cc}$ and $\mathcal{S}_{cc}$ from Table~\ref{tab:Coeffs:dhbaryon-lattice},
we get the prediction $(3585 \pm 12)$~MeV, which is about 3  error bars below the measured mass.

The diquark energy $\mathcal{E}_{Q_1Q_2}$ in the doubly heavy baryon can be estimated 
using the energy $\mathcal{E}_{u/d,Q_1Q_2}$  in Table~\ref{tab:Coeffs:dhbaryon-lattice} 
and the energies $\mathcal{E}_{u/d,Q}$ in heavy mesons from Table~\ref{tab:Coeffs:hmeson-PDG}.
These energies are defined in Eqs.~\eqref{E_ell,Q1Q2} and \eqref{E_ell,Q}, respectively.
The energies of the $cc$, $bb$, and  $bc$ diquarks can be expressed up to errors of order $1/m_Q^2$ as
%%%%%%%%
\begin{subequations}
\begin{eqnarray}
\mathcal{E}_{cc} &=& \mathcal{E}_{u/d,cc} - \frac{m_b-2m_c}{2(m_b - m_c)}  \mathcal{E}_{u/d,c} 
- \frac{m_b}{2(m_b - m_c)} \mathcal{E}_{u/d,b},
\label{Ecc}
\\
\mathcal{E}_{bb} &=& \mathcal{E}_{u/d,bb} + \frac{m_c}{2(m_b - m_c)} \mathcal{E}_{u/d,c} 
- \frac{2m_b- m_c}{2(m_b - m_c)} \mathcal{E}_{u/d,b} ,
\label{Ebb}
\\
\mathcal{E}_{bc} &=& \mathcal{E}_{u/d,bc} 
 + \frac{m_c^2}{m_b^2-m_c^2}  \mathcal{E}_{u/d,c}  - \frac{m_b^2}{m_b^2-m_c^2}  \mathcal{E}_{u/d,b}.
\label{Ebc}
\end{eqnarray}
\end{subequations}
%%%%%%%%
The coefficients of $\mathcal{E}_{u/d}$ and $\mathcal{K}_{u/d}$ on the right sides cancel.
The predicted energy of the $cc$ diquark is $\mathcal{E}_{cc} = (11 \pm 11)$~MeV.
The predicted  energy of the $bb$ diquark is $\mathcal{E}_{bb} = (-153 \pm 25)$~MeV.
The predicted energy of a $bc$ diquark is $\mathcal{E}_{[bc]} = (-30 \pm 37)$~MeV
in the case of antisymmetric heavy flavor and $\mathcal{E}_{\{ bc\}} = (4 \pm 27)$~MeV
in the case of symmetric heavy flavor.
A sufficiently negative value of $\mathcal{E}_{Q_1Q_2}$ implies the binding of the two heavy quarks into a diquark.
The $bb$ diquark is the only one for which our estimate suggests a significant binding energy. 

It is interesting to compare lattice QCD results for hyperfine splittings 
with the predictions of  heavy-quark--diquark symmetry.
The prediction for the hyperfine splitting between $\Xi_{cc}^{*++}$ and $\Xi_{cc}^{++}$ 
from heavy-quark--diquark symmetry in Eq.~\eqref{M*-M:HQD} is 107~MeV.  
The result from lattice QCD in Ref.~\cite{Brown:2014ena}, with statistical and systematic errors added linearly, 
is $83 \pm 13$~MeV, which is lower by about 2 error bars.
The prediction for the hyperfine splitting between $\Xi_{bb}^{*++}$ and $\Xi_{bb}^{++}$ 
from heavy-quark--diquark symmetry analogous to Eq.~\eqref{M*-M:HQD} is 34~MeV.  
The result from lattice QCD in Ref.~\cite{Brown:2014ena}, with statistical and systematic errors added linearly, 
is $35 \pm 10$~MeV, which is consistent within the error.
Our expression for the spin-splitting coefficient $\mathcal{S}_{\ell, bc}$ for $bc$ baryons in Eq.~\eqref{S,bc}
implies the inequalities
$\mathcal{S}_{\ell, \bar b} < \mathcal{S}_{\ell, bc} < \mathcal{S}_{\ell, \bar c}$.
They imply corresponding inequalities between the hyperfine splittings of $b$ mesons, $bc$ baryons,  and $c$ mesons.
They imply, for example,
%%%%%%%%
\begin{equation}
\frac34 \big( M_{B^{*+}} -  M_{B^+} \big) < M_{\Xi_{bc}^{*+}} -  M_{\Xi_{bc}^{\prime +}} 
<   \frac34 \big( M_{D^{*0}} -  M_{D^0} \big).
\label{M*-M:b<bc<c}
\end{equation}
%%%%%%%%
The hyperfine splitting between $\Xi_{bc}^{*+}$ and $\Xi_{bc}^{\prime +}$ is predicted to be between
34~MeV and 107~MeV.
The hyperfine splitting calculated using lattice QCD in Ref.~\cite{Brown:2014ena},
with statistical and systematic errors added linearly,  is  $27 \pm 12$~MeV.
The central value is below the lower bound in Eq.~\eqref{M*-M:b<bc<c} but by less than an error bar.
If the charm quark is treated as a light quark instead of as a constituent of the $bc$ diquark,
the prediction for the spin-splitting coefficient is $\mathcal{S}_{\ell, bc} = \mathcal{S}_{\ell’, b}$,
where $\mathcal{S}_{\ell’, b}$ is the spin-splitting coefficient for the ground-state $b$ baryons.
The corresponding prediction for the hyperfine splitting between $\Xi_{bc}^{*+}$ and $\Xi_{bc}^{\prime +}$ is
%%%%%%%%
\begin{equation}
M_{\Xi_{bc}^{*+}} -  M_{\Xi_{bc}^{\prime +}} 
= M_{\Xi_b^{*-}} -  M_{\Xi_b^{\prime -}} .
\label{M*-M:bc}
\end{equation}
%%%%%%%%
This hyperfine splitting is predicted to be 20~MeV,
which is consistent with  the  lattice QCD prediction $27 \pm 12$~MeV.

%%%%%%%%%%%%%%%%%%%%%%%%%%%%%%%%%%%%%%%%%%%%%%%
\subsection{Predictions for Doubly Heavy Tetraquarks}
\label{sec:2Tetraquark}
%%%%%%%%%%%%%%%%%%%%%%%%%%%%%%%%%%%%%%%%%%%%%%%

The masses of the ground-state doubly heavy tetraquarks
can be predicted using the Hamiltonian in Eq.~\eqref{H_ell,Q1Q2}.
Our choices for the heavy quark masses $m_c$ and $m_b$ are given in Eqs.~\eqref{mQ}.
We determine the coefficients in the Hamiltonian for doubly heavy tetraquarks 
in Eq.~\eqref{H_ell,Q1Q2} 
from the coefficients for doubly heavy baryons in Table~\ref{tab:Coeffs:dhbaryon-lattice},
the coefficients for heavy baryons in Table~\ref{tab:Coeffs:hbaryon-PDG}, and the coefficients for heavy mesons in Table~\ref{tab:Coeffs:hmeson-PDG}.
The energies $\mathcal{E}_{u/d,Q_1Q_2}$ for doubly heavy tetraquarks 
in Table~\ref{tab:Coeffs:dhtetraquark}
are determined from  $\mathcal{E}_{u/d,Q_1Q_2}$ for doubly heavy baryons in Table~\ref{tab:Coeffs:dhbaryon-lattice},
$\mathcal{E}_{u/d,c}$ and $\mathcal{E}_{u/d,b}$ for heavy baryons in Table~\ref{tab:Coeffs:hbaryon-PDG},
and   $\mathcal{E}_{u/d,c}$ and $\mathcal{E}_{u/d,b}$ for  heavy mesons in Table~\ref{tab:Coeffs:hmeson-PDG}
by using Eqs.~\eqref{E_ell,cc,bb,bc}.
The strange-quark masses    $m_{s,Q_1Q_2}$ for doubly heavy tetraquarks in Table~\ref{tab:Coeffs:dhtetraquark} 
are  determined from the strange-quark masses  $m_{s,c}$ and $m_{s,b}$ for heavy
baryons in Table~\ref{tab:Coeffs:hbaryon-PDG}  by using Eqs.~\eqref{ms,cc,bb,bc}.
The spin-splitting coefficients
$\mathcal{S}_{Q_1Q_2}$ for doubly heavy tetraquarks in Table~\ref{tab:Coeffs:dhtetraquark} 
are determined from the coefficients
$\mathcal{S}_b$ and $\mathcal{S}_c$ for heavy baryons in Table~\ref{tab:Coeffs:hbaryon-PDG}
using Eqs.~\eqref{S,QQ} and Eq.~\eqref{S,bc}.
The results are all given  in  Table~\ref{tab:Coeffs:dhtetraquark}.

%%%%%%%%%%%%%%%%%%%%%%%%%%%%%%%%%%%%%%%%%%
\begin{table}[t]
\begin{center}
\begin{tabular}{cc|cccl}
%\rule[10pt]{-1mm}{0mm}
~$Q_1Q_2$~&~$\ell$~&~$\mathcal{E}_{u/d,Q_1Q_2}$ [MeV]~&~$m_{s,Q_1Q_2}$ [MeV]~&
~$\mathcal{S}_{Q_1Q_2}$ [GeV$^2$]~ & \\
\hline
$c\, c$      & $[\bar q\, \bar q']$, $0^+$ & $627.4 \pm 11.2$ & $176.8 \pm 2.9$ &   \\
$[b\, c]$    & $[\bar q\, \bar q']$, $0^+$ & $581.0 \pm 37.4$ & $173.8 \pm 4.3$ &                             \\
$\{b\, c\}$ & $[\bar q\, \bar q']$, $0^+$ & $614.5 \pm 27.5$ &  $173.8 \pm 4.3$ &  \\
$b\, b$     & $[\bar q\, \bar q']$, $0^+$ & $456.4 \pm 25.4$ &  $172.8 \pm 4.8$ &  \\
\hline
$c\, c$      & $\{\bar q\, \bar q'\}$, $1^+$ & $835.7 \pm 11.1$ & $120.2 \pm 1.7$ & $0.147 \pm 0.003$ \\
$[b\, c]$    & $\{\bar q\, \bar q'\}$, $1^+$ & $788.0 \pm 37.3$ & $118.1 \pm 2.5$ &                                \\
$\{b\, c\}$ & $\{\bar q\, \bar q'\}$, $1^+$ & $821.5 \pm 27.4$ &  $118.1 \pm 2.5$ & $0.145 \pm 0.006$ & \\
$b\, b$     & $\{\bar q\, \bar q'\}$, $1^+$ & $662.8 \pm 25.2$ &  $117.5 \pm 2.8$ & $0.137 \pm 0.024$ & \\
 \end{tabular}
\end{center}
\caption{
Coefficients in the Hamiltonian in Eq.~\eqref{H_ell,Q1Q2} for ground-state doubly heavy tetraquarks. 
The energies  $\mathcal{E}_{u/d,Q_1Q_2}$ and the strange-quark masses $m_{s,Q_1Q_2}$ 
are obtained from those in Tables~\ref{tab:Coeffs:hmeson-PDG}, \ref{tab:Coeffs:hbaryon-PDG}, 
and \ref{tab:Coeffs:dhbaryon-lattice} using  
Eqs.~\eqref{E_ell,cc,bb,bc} and \eqref{ms,cc,bb,bc}.
The  spin-splitting  coefficients $S_{cc}$  and  $S_{bb}$ are 
equal to $S_c$ and $S_b$ in Table~\ref{tab:Coeffs:hbaryon-PDG},
while $S_{\{b\, c\}}$ is the linear combination of $S_c$ and $S_b$ in Eq.~\eqref{S,bc}.
}
\label{tab:Coeffs:dhtetraquark}
\end{table}
%%%%%%%%%%%%%%%%%%%%%%%%%%%%%%%%%%%%%%%%%%

The masses of the ground-state doubly heavy tetraquarks can be predicted by inserting the coefficients in
Table~\ref{tab:Coeffs:dhtetraquark} into the Hamiltonian in  Eq.~\eqref{H_ell,Q1Q2}.
The resulting predictions for the masses of $cc$ and $bb$ tetraquarks are given in 
Table~\ref{tab:cc_bbTetraquarks}. 
The resulting predictions for the masses of $bc$ tetraquarks are given in Table~\ref{tab:bcTetraquarks}.
The strong-decay threshold  in the last column of Tables~\ref{tab:cc_bbTetraquarks} or \ref{tab:bcTetraquarks} 
is the sum of the masses of the lightest pair of heavy mesons into which the tetraquark can decay.
The only tetraquarks with masses below the strong-decay thresholds are
the one with  flavor $bb[\bar u \bar d]$, which is below the threshold by $(133 \pm 25)$~MeV,
and those with flavor $bb[\bar s \bar u]$ or $bb[\bar s \bar d]$, which are below the threshold by $(48 \pm 26)$~MeV.
The $cc$ tetraquark with mass closest to the strong-decay threshold has flavor  $cc[\bar u \bar d]$,
and it is above the threshold by $72 \pm 11$~MeV.
The $bc$ tetraquarks with masses closest to the strong-decay threshold have flavors $[bc][\bar u \bar d]$
and $\{bc\}  [\bar u \bar d]$, and they are above the threshold by $104 \pm 37$~MeV and  $92 \pm 28$~MeV, respectively.

%%%%%%%%%%%%%%%%%%%%%%%%%%%%%%%%%%%%%%%%%%
\begin{table}[ht!]
\begin{center}
\begin{tabular}{lc|ll|ll}
%\rule[10pt]{-1mm}{0mm}
flavor                          &         $J^P$        &~Eichten-Quigg    &~this work~       &  ~threshold~  \\
\hline
$cc [\bar u \bar d]$     &         $1^+$        &~3978                   &~$3947 \pm 11$                      &~3875                &\\
$cc [\bar s \bar u]$     &         $1^+$        &~4156                   &~$4124\pm 12$                       &~3975                &\\
$cc \{\bar u \bar d\}$~&~$0^+,1^+,2^+$~&~$4146+(0,21,64)$~&~$4111+(0,22,66)\pm 11$~&~$3734+(0,141,0)$~&\\
$cc \{\bar s \bar u\}$  &  $0^+,1^+,2^+$  &                             &~$4232+(0,22,66)\pm 11$  &~$3833+(0,142,0)$~&\\
$cc \, \bar s \bar s$    &  $0^+,1^+,2^+$  &                             &~$4352+(0,22,66)\pm 12$  &~$3937+(0,144,0)$~&\\
\hline
$bb [\bar u \bar d]$        &~$1^+$               &~$\bm{10482}$     &~$\bm{10471 \pm 25}$ &~10604                  & \\
$bb [\bar s \bar u]$        &~$1^+$               &~$\bm{10643}$     &~$\bm{10644 \pm 26}$ &~10692                  &\\
$bb \{\bar u \bar d\}$     &~$0^+,1^+,2^+$  &~$10674+(0,7,21)$~&~$10664+(0,7,21) \pm 25$~&~$10559+(0,45,0)$~  & \\
$bb \{\bar s \bar u\}$     &~$0^+,1^+,2^+$  &                             &~$10781+(0,7,21) \pm 25$  &~$10646+(0,45,0)$     & \\
 $bb\, \{\bar s \bar s\}$~&~$0^+,1^+,2^+$~&                             &~$10898+(0,7,21) \pm 26$  &~$10734+(0,49,0)$     &\\
 \end{tabular}
\end{center}
\caption{Predicted masses of ground-state $c c$ and $bb$ tetraquarks in the heavy-diquark limit. 
Only one member of any isospin multiplet is given.
All masses are in MeV.
The results labeled ``Eichten-Quigg''  are from Ref.~\cite{Eichten:2017ffp}.
The results labeled ``this work'' were obtained using the Hamiltonian in Eq.~\eqref{H_ell,Q1Q2}
with the coefficients in Table~\ref{tab:Coeffs:dhtetraquark}.
The last column is the strong-decay threshold. The bold-faced energies are below the strong-decay threshold.}
\label{tab:cc_bbTetraquarks}
\end{table}
%%%%%%%%%%%%%%%%%%%%%%%%%%%%%%%%%%%%%%%%%%

%%%%%%%%%%%%%%%%%%%%%%%%%%%%%%%%%%%%%%%%%%
\begin{table}[ht!]
\begin{center}
\begin{tabular}{lc|ll|ll}
%\rule[10pt]{-1mm}{0mm}
flavor                              &          $J^P$        &~Eichten-Quigg  &~this work~                          &~threshold~  \\
\hline
$[bc] [\bar u \bar d]$       &          $0^+$        &~7229                 &~$7248 \pm 37$                  &~7144                &\\
$[bc] [\bar s \bar u]$       &          $0^+$        &~7406                 &~$7422 \pm 38$                  &~7232                &\\
$[bc] \{\bar u \bar d\}$     &         $1^+$         &~7439                 &~$7455 \pm 37$                  &~7190                &\\
$[bc] \{\bar s \bar u\}$     &         $1^+$         &                           &~$7573 \pm 37$                  &~7280                &\\
$[bc]\, \bar s \bar s$        &         $1^+$         &                           &~$7691 \pm 38$                  &~7384                &\\
\hline
$\{bc\} [\bar u \bar d]$     &         $1^+$         &~7272                  &~$7282 \pm 28$                 &~7190                &\\
$\{bc\} [\bar s \bar u]$     &         $1^+$         &~7445                  &~$7456 \pm 28$                 &~7280                &\\
$\{bc\} \{\bar u \bar d\}$~&~$0^+,1^+,2^+$~&~$7461+(0,11,32)$~&~$7460+(0,14,43)\pm 27$~&~$7144+(0,45,0)$~&
\\
$\{bc\} \{\bar s \bar u\}$  &  $0^+,1^+,2^+$  &                            &~$7578+(0,14,43)\pm 28$   &~$7232+(0,49,0)$   &
\\
$\{bc\} \,\bar s \bar s$     &  $0^+,1^+,2^+$  &                            &~$7696+(0,14,43)\pm 28$   &~$7335+(0,49,0)$  &
\\
 \end{tabular}
\end{center}
\caption{Predicted masses of ground-state $bc$ tetraquarks in the heavy-diquark limit. 
Only one member of any isospin multiplet is given.
All masses are in MeV.
The results labeled ``Eichten-Quigg''  are from Ref.~\cite{Eichten:2017ffp}.
The results labeled ``this work'' were obtained using the Hamiltonian in Eq.~\eqref{H_ell,Q1Q2}
with the coefficients in Table~\ref{tab:Coeffs:dhtetraquark}.
The last column is the strong-decay threshold. 
}
\label{tab:bcTetraquarks}
\end{table}
%%%%%%%%%%%%%%%%%%%%%%%%%%%%%%%%%%%%%%%%%%

The predictions for the masses of doubly heavy tetraquarks by Eichten and Quigg (EQ) in Ref.~\cite{Eichten:2017ffp}
are also given in Tables~\ref{tab:cc_bbTetraquarks} and \ref{tab:bcTetraquarks}.
They did not give any error bars on their predictions.
The masses of doubly heavy baryons they used as inputs were the measured mass of $\Xi_{cc}^{++}$ 
and the masses of other doubly heavy baryons predicted by the quark model in Ref.~\cite{Karliner:2014gca}.
The predictions of EQ for the masses of $cc$ tetraquarks are higher than ours by about 30~MeV, 
which is about 3 error bars.
The predictions of EQ  for the masses of  $bb$ tetraquarks and $bc$ tetraquarks
agree with our predictions to within our error bars.

In Ref.~\cite{Karliner:2017qjm}, Karliner and Rosner (KR) used  a quark-diquark model
to predict the masses of the most deeply bound doubly heavy tetraquarks.
They gave  error bars on their predictions of approximately $\pm 12$~MeV for every mass.
Those error bars  were not based on any serious error analysis.
Their prediction for the mass of the $bb [\bar u \bar d]$ tetraquark was $(10389 \pm 12)$~MeV,
which is 215~MeV below the strong-decay threshold.  Our prediction in  Table~\ref{tab:cc_bbTetraquarks}
is higher than theirs by 82~MeV, which is 3.3 of our error bars. 
Their prediction for the mass of the $[bc] [\bar u \bar d]$  tetraquark was $(7134 \pm 13)$~MeV,
which is only 10~MeV below the strong-decay threshold. Our   prediction in Table~\ref{tab:bcTetraquarks}
is higher than theirs by 114~MeV, which is  about 3 of our error bars. 
Their prediction for the mass of the $cc [\bar u \bar d]$ tetraquark was $(3882 \pm 12)$~MeV,
which is 7~MeV above the strong-decay threshold.  Our  prediction in Table~\ref{tab:cc_bbTetraquarks}
is higher than theirs by 65~MeV, which is about 6 of our error bars.

There have been several calculations of the masses of doubly heavy tetraquarks using lattice QCD.
Francis {\it et al.}\ presented strong evidence for the existence of deeply bound tetraquark states 
with flavors $bb\bar u \bar d$,  $bb\bar s \bar u$, and $bb\bar s \bar d$ \cite{Francis:2016hui}.
Their simulations were extrapolated to the physical values of the light quark masses,
but they were carried out at a single lattice spacing and volume.
They used the correlators of two local operators to determine the masses.
They also provided evidence for the  existence of bound tetraquark states 
with flavor $bc\bar u \bar d$  \cite{Francis:2018jyb}.
Junnarkar, Mathur and Padmanath verified the existence of deeply bound tetraquark states 
with flavors $bb\bar u \bar d$ and  $bb\bar s \bar u$ \cite{Junnarkar:2018twb}.
They also found that the $cc\bar u \bar d$ and  $cc\bar s \bar u$ tetraquark states were below 
but close to the relevant meson pair thresholds.
Their simulations were carried out at three lattice spacings but only at a single volume,
and they were extrapolated to the physical values of the light quark masses.
They used the correlators of two local operators to determine the masses.
Leskovec {\it et al.}\ calculated the mass of the ground-state $bb\bar u \bar d$ tetraquark  with quantum numbers $1^+$
and quantified all the major systematic errors \cite{Leskovec:2019ioa}.
Their simulations were carried out using five lattice gauge ensembles with various lattice spacings 
and various light quark masses, including the physical values,
and they were extrapolated to infinite volume.
They used the correlators of three local operators and two bilocal operators to determine the mass.
Their result for the mass is $10476 \pm 24 \pm 10$~MeV.
This result is in excellent agreement with our  prediction for the $bb [\bar u \bar d]$ tetraquark
in Table~\ref{tab:cc_bbTetraquarks}, which has comparable errors.

%\newpage

%%%%%%%%%%%%%%%%%%%%%%%%%%%%%%%%%%%%%%%%%%%%%%%
\section{Discussion}
\label{sec:Discussion}
%%%%%%%%%%%%%%%%%%%%%%%%%%%%%%%%%%%%%%%%%%%%%%%

We have presented predictions for the masses of doubly heavy tetraquarks with error bars.
We followed the general strategy  that Eichten and Quigg used to provide convincing evidence 
that there are stable $bb$ tetraquarks  with masses below their strong-decay thresholds \cite{Eichten:2017ffp}.
Our analysis was based on the Hamiltonian for doubly heavy hadrons in Eq.~\eqref{HQ1Q2_ell},
which provides an expansion for their masses to first order in the inverse heavy quark masses $1/m_Q$.
The analogous Hamiltonian for singly heavy hadrons is given in Eq.~\eqref{H_ell}.
The approximate heavy-quark--diquark symmetry of QCD relates the coefficients in the Hamiltonians
for doubly heavy baryons and doubly heavy tetraquarks to those for heavy mesons and heavy baryons, respectively.
We allowed for dependence of the coefficients of the spin-dependent terms in the Hamiltonians
on the heavy flavors.  The resulting Hamiltonians for heavy hadrons and for doubly heavy hadrons 
are given in Eqs.~\eqref{H_ell,Q} and \eqref{H_ell,Q1Q2}, respectively.
The coefficients in the Hamiltonians for the ground-state heavy mesons and for the ground-state heavy  baryons 
were determined from measured hadron masses and are given in Tables~\ref{tab:Coeffs:hmeson-PDG} 
and \ref{tab:Coeffs:hbaryon-PDG}, respectively.
The coefficients in the Hamiltonian for the ground-state doubly heavy  baryons were determined 
from baryon masses calculated using lattice QCD and are given in Table~\ref{tab:Coeffs:dhbaryon-lattice}.
The coefficients in Tables~\ref{tab:Coeffs:hmeson-PDG}, \ref{tab:Coeffs:hbaryon-PDG}, 
and \ref{tab:Coeffs:dhbaryon-lattice} were used to determine the coefficients in the Hamiltonian 
for the ground-state doubly heavy tetraquarks, which are given with error bars in Table~\ref{tab:Coeffs:dhtetraquark}.
Those coefficients were then used to predict the masses of the ground-state $cc$ and $bb$ tetraquarks
in Table~\ref{tab:cc_bbTetraquarks} and  the masses of the ground-state $bc$ tetraquarks
in Table~\ref{tab:bcTetraquarks}.

 Our Hamiltonian for doubly heavy hadrons in Eq.~\eqref{HQ1Q2_ell} was deduced from the Lagrangian 
 for an effective field theory for doubly heavy hadrons developed by Brambilla, Vairo, and Rosch (BVR) 
 \cite{Brambilla:2005yk} and by Fleming and Mehen \cite{Fleming:2005pd}.
The terms of order $1/m_Q$ in the BVR Lagrangian that involve the triplet diquark field $T$
and light QCD fields were used to deduce corresponding terms in the Hamiltonian in Eq.~\eqref{HQ1Q2_ell}.
We ignored the   terms  in the  BVR Lagrangian  that involve the sextet diquark field $\Sigma$.
The coefficients  in the Hamiltonian for doubly heavy hadrons  in Eq.~\eqref{HQ1Q2_ell}
have factors $\mathcal{E}_\ell$, $\mathcal{K}_\ell$, and $\mathcal{S}_\ell$ 
that depend on the discrete state $\ell$ of the light QCD fields.
Those same factors appear  in the  Hamiltonian for singly heavy hadrons.
The dependence of the coefficients on the heavy quark masses $m_c$ and $m_b$ 
 is determined by  the BVR Lagrangian.

Our analysis differs from that of Eichten and Quigg (EQ) in Ref.~\cite{Eichten:2017ffp} in several important ways.
Instead of setting the heavy quark masses $m_c$ and $m_b$ equal to half the masses of the  
quarkonium states $J/\psi$ and $\Upsilon$,  respectively, we set them equal to the linear combinations 
of the heavy meson masses and the heavy baryon mass in Eqs.~\eqref{mQ}.
We corrected
an error in the coefficient of the spin-dependent term in the Hamiltonian for doubly heavy hadrons
in Ref.~\cite{Eichten:2017ffp}.  
EQ assumed the denominator of the coefficient of the spin-dependent term in the Hamiltonian  
was proportional to $m_{Q_1}+m_{Q_2}$ instead of the reduced mass $\mu_{Q_1Q_2}$.
Their coefficient is fortuitously correct in the case of identical heavy quarks,
but the correct coefficient is larger by a factor of 1.34 in the case of $bc$ hadrons. 
We also improved upon the analysis of EQ by using 
the masses of doubly heavy baryons calculated using lattice QCD as inputs
 instead of the predictions of the quark model in Ref.~\cite{Karliner:2014gca}.
 Since all of our inputs have well defined  error bars, we were able to predict the 
 masses of tetraquarks with error bars.

Our results are in surprisingly good agreement with those of Eichten and Quigg \cite{Eichten:2017ffp}.
Their predictions for the masses of $bb$ tetraquarks and $bc$ tetraquarks agree with ours to within our errors, 
but their predictions for the masses of $cc$ tetraquarks  are higher by about 30~MeV.
We agree with Eichten and Quigg that the only doubly heavy  tetraquarks  with masses below the strong-decay thresholds
are the ground states with flavor $bb[\bar u \bar d]$, $bb[\bar s \bar u]$, and $bb[\bar s \bar d]$.
Our prediction for the mass of the $bb[\bar u \bar d]$ tetraquark is  in very good agreement
with the lattice QCD calculation in Ref.~\cite{Leskovec:2019ioa}.
Our prediction for the mass is larger than that of Karliner and Rosner \cite{Karliner:2017qjm}
by 3.3 of our error bars.   Our predictions for the masses of
the ground-state $bc$ and $cc$ tetraquarks are  well above their strong-decay thresholds, in 
 agreement with the results of Eichten and Quigg and
contrary to the predictions of Karliner and Rosner \cite{Karliner:2017qjm}.

Our predictions for the masses of doubly heavy tetraquarks could be made more precise
by additional calculations of the masses of doubly heavy baryons using lattice QCD.
Additional calculations of the masses of $bc$ and $bb$ baryons would be especially helpful.
The only such calculations in which all the important systematic errors have been quantified
are the pioneering calculations by Brown {\it et al.}\ \cite{Brown:2014ena}.
In order for the lattice QCD calculations to provide the greatest possible insights, 
it is important that results are given not only for individual masses, 
but also for appropriate mass differences, as in Ref.~\cite{Brown:2014ena}.
The error bars in hyperfine splittings can be much smaller than those obtained by subtracting the masses.

Our analysis of doubly heavy baryons using masses calculated using lattice QCD raises questions 
about the accuracy of heavy-quark--diquark symmetry for doubly heavy hadrons containing charm quarks.
The spin-splitting coefficient $\mathcal{S}_{bb}$ for $bb$ baryons in Table~\ref{tab:Coeffs:dhbaryon-lattice} 
is compatible within errors with $\mathcal{S}_b$ for bottom mesons in Table~\ref{tab:Coeffs:hmeson-PDG}.
However $\mathcal{S}_{cc}$ for $cc$ baryons in Table~\ref{tab:Coeffs:dhbaryon-lattice} 
is smaller than $\mathcal{S}_c$ for charm mesons in Table~\ref{tab:Coeffs:hmeson-PDG} by several error bars.
A more striking problem is that $\mathcal{S}_{\{ b c \}}$ for $bc$ baryons in Table~\ref{tab:Coeffs:dhbaryon-lattice} 
is significantly smaller than either $\mathcal{S}_{cc}$ or $\mathcal{S}_{bb}$,
instead of being intermediate between $\mathcal{S}_{cc}$ and $\mathcal{S}_{bb}$ as predicted by Eq.~\eqref{S,bc}.
The spin-splitting coefficient $\mathcal{S}_{\{ b c \}}$ is however  compatible within errors 
with $\mathcal{S}_b$ for bottom baryons in Table~\ref{tab:Coeffs:hbaryon-PDG}.
This suggests that it may be a better approximation to treat the charm quark in a $bc$ hadron as a light quark
instead of as a constituent of a $bc$ diquark.
In this case, $bc$ baryons and $bc$ tetraquarks would not be related by heavy-quark--diquark symmetry.

Our analysis was based on the assumption that the two heavy quarks in a doubly heavy hadron 
form a compact diquark in a $\bf{3^{*}}$ color state. The diquark can also be in a {\bf 6} color  state.
In the effective field theory of Ref.~\cite{Brambilla:2005yk}, 
the color-sextet component of the heavy diquark is taken into account through terms in the BVR Lagrangian 
that involve the sextet field $\Sigma$.
If the effects of $\Sigma$ are significant,
the Hamiltonian for doubly heavy hadrons would not have the simple form in Eq.~\eqref{H_ell,Q1Q2},
with coefficients related to those in the Hamiltonian for  heavy hadrons by heavy-quark--diquark symmetry.

The two heavy quarks in a doubly heavy hadron can be treated as a diquark only if their separation 
is smaller than the  length scale of the light QCD fields.
It may be possible to take into account contributions from heavy quarks with larger separations 
using the Born-Oppenheimer approximation.
The Born-Oppenheimer approximation  for QCD was pioneered by Juge, Kuti, and Morningstar,
who applied it to heavy quarkonium and to heavy quarkonium hybrids \cite{Juge:1999ie}.
The application of the Born-Oppenheimer approximation to exotic heavy hadrons 
that contain a $Q \bar Q$ pair and a light quark-antiquark pair was proposed in Ref.~\cite{Braaten:2014qka}.
Brambilla {\it et al.}\ have developed an effective field theory framework 
based on the Born-Oppenheimer approximation \cite{Brambilla:2017uyf}.
Such a framework has been applied extensively to heavy quarkonium hybrids 
\cite{Berwein:2015vca,Oncala:2017hop,Brambilla:2018pyn,Brambilla:2019jfi}.
The Born-Oppenheimer approximation  for $QQ$ hadrons was pioneered by Bicudo  {\it et al.},
who applied it to doubly heavy tetraquarks \cite{Bicudo:2012qt}.
Bicudo {\it et al.}\  used lattice QCD calculations of the static potentials for two heavy quarks
together with the Born-Oppenheimer approximation to present evidence for the existence 
of stable $bb$  tetraquarks \cite{Bicudo:2015vta,Bicudo:2016ooe}.

The Born-Oppenheimer approximation suggests that deviations from the  compact heavy diquark limit
 are very different for doubly heavy tetraquarks and doubly heavy baryons.  
The difference arises because the constituents of a doubly heavy tetraquark
can be rearranged into two color-singlet clusters, while those of a doubly heavy baryon cannot.
Forming two color-singlet clusters from the constituents of a doubly heavy baryon 
requires the creation of a quark-antiquark pair, 
and this is a process that is dynamically suppressed in low-energy QCD. 
The  Born-Oppenheimer potentials between the two heavy quarks in a tetraquark and in a baryon 
therefore differ dramatically as their separation increases.
For doubly heavy tetraquarks, the Born-Oppenheimer potentials display screening behavior,
approaching constants near the thresholds for pairs of heavy mesons.
For doubly heavy baryons, the  Born-Oppenheimer potentials instead display string-breaking behavior,
with avoided crossings near the thresholds for a heavy meson and a heavy baryon.

A new nonrelativistic effective field theory for doubly heavy hadrons beyond the  compact heavy-diquark limit  was recently developed by Soto and Tarr\'us Castell\`a \cite{Soto:2020xpm}
and applied to doubly heavy baryons \cite{Soto:2020pfa}. 
At leading order in the $1/m_Q$ expansion, the  Lagrangian 
has heavy quark spin  symmetry and corresponds to the Born-Oppenheimer approximation. 
At next-to-leading order in the $1/m_Q$ expansion, the  Lagrangian 
includes terms that depend on the total spin and the orbital angular momentum of the two heavy quarks.  
Soto and Tarr\'us Castell\`a calculated the spectra of  $cc$ and $bb$ baryons 
at leading order in the $1/m_Q$ expansion
by solving  a Schroedinger equation with three coupled channels \cite{Soto:2020pfa} using
Born-Oppenheimer  potentials, extracted from lattice QCD calculations 
with unphysically large light-quark masses.  They found that the compact heavy-diquark limit was not a good approximation
for orbital-angular-momentum excitations of the heavy quark pair, even in the case of the $b$ quark.
Using the measured mass of the $\Xi_{cc}^{++}$ as input \cite{Aaij:2017ueg,Aaij:2018gfl},
they predicted the hyperfine splitting between the $\Xi_{cc}^{*++}$ and $\Xi_{cc}^{++}$ to be $136 \pm 44$~MeV.
This is consistent to within the errors with the prediction of 107~MeV from heavy-quark--diquark symmetry 
in Eq.~\eqref{M*-M:HQD}.

While lattice QCD will provide definitive calculations of the masses of some
doubly heavy tetraquarks, a more complete picture of their spectra can be
obtained by using lattice QCD in conjunction with other theoretical methods.
We have used lattice QCD calculations of masses of doubly heavy baryons 
in conjunction with constraints deduced from the effective field theory pNRQCD  for doubly heavy hadrons 
to predict the masses of doubly heavy tetraquarks with error bars. 
A crucial assumption in our analysis is  that the two heavy quarks form a compact diquark.
That assumption is avoided in the effective field theory for doubly heavy
hadrons developed in Refs.~\cite{Soto:2020pfa,Soto:2020xpm}, which makes use of lattice QCD calculations of Born-Oppenheimer potentials.  Such an effective field theory, along with the 
discovery of more doubly heavy hadrons in future experiments, 
should eventually provide a complete picture of the spectrum of doubly heavy tetraquarks.

\begin{acknowledgments}
This research was supported in part by the U.S.\ Department of Energy under grant DE-FG02-05ER15715
%and by Director, Office of Science, Office of Nuclear Physics, of the U.S.\ Department of Energy under 
and grant DE-FG02-05ER41368.
EB would like to thank E.~Eichten for valuable discussions.
This work was initiated during a program at the Munich Institute for Astro- and Particle Physics
 in October 2019. 
EB thanks Prof. N. Brambilla for arranging a visiting professorship at the Technical University of Munich
supported by the Bavarian State Ministry of Education, Science, and Arts.

\end{acknowledgments}

%\newpage

%\appendix

%\newpage

%%%%%%%%%%%%%%%%%%%%%%%%%%%%%%%%%%%%%%%%%%%%%%%

%%%%%%%%%%%%%%%%%%%%%%%%%%%%%%%%%%%%%%%%%%%%%%%


\begin{thebibliography}{99}
%%%%%%%%%%%%%%%%%%%%%%%%%%%%%%%%%%%%%%%%%%%%%%%

%\cite{Ader:1981db}
\bibitem{Ader:1981db} 
  J.P.~Ader, J.M.~Richard and P.~Taxil,
Do Narrow Heavy Multi-Quark States Exist?,
  Phys.\ Rev.\ D {\bf 25}, 2370 (1982).
%  doi:10.1103/PhysRevD.25.2370
  %%CITATION = doi:10.1103/PhysRevD.25.2370;%%
  
%\cite{Zouzou:1986qh}
\bibitem{Zouzou:1986qh}
S.~Zouzou, B.~Silvestre-Brac, C.~Gignoux and J.~Richard,
%``FOUR QUARK BOUND STATES,''
Four-quark bound states,
Z.\ Phys. C \textbf{30}, 457 (1986).
%doi:10.1007/BF01557611


%\cite{Manohar:1992nd}
\bibitem{Manohar:1992nd} 
  A.V.~Manohar and M.B.~Wise,
Exotic $Q Q \bar q \bar q$ states in QCD,
  Nucl.\ Phys.\ B {\bf 399}, 17 (1993)
%  doi:10.1016/0550-3213(93)90614-U
  [hep-ph/9212236].
  %%CITATION = doi:10.1016/0550-3213(93)90614-U;%%
  
%\cite{Savage:1990di}
\bibitem{Savage:1990di} 
  M.J.~Savage and M.B.~Wise,
Spectrum of baryons with two heavy quarks,
  Phys.\ Lett.\ B {\bf 248}, 177 (1990).
 % doi:10.1016/0370-2693(90)90035-5
  %%CITATION = doi:10.1016/0370-2693(90)90035-5;%%
  
%\cite{Brambilla:2005yk}
\bibitem{Brambilla:2005yk} 
  N.~Brambilla, A.~Vairo and T.~Rosch,
Effective field theory Lagrangians for baryons with two and three heavy quarks,
  Phys.\ Rev.\ D {\bf 72}, 034021 (2005)
%  doi:10.1103/PhysRevD.72.034021
  [hep-ph/0506065].
  %%CITATION = doi:10.1103/PhysRevD.72.034021;%%  
  
%\cite{Fleming:2005pd}
\bibitem{Fleming:2005pd} 
  S.~Fleming and T.~Mehen,
Doubly heavy baryons, heavy quark-diquark symmetry and NRQCD,
  Phys.\ Rev.\ D {\bf 73}, 034502 (2006)
%  doi:10.1103/PhysRevD.73.034502
  [hep-ph/0509313].
  %%CITATION = doi:10.1103/PhysRevD.73.034502;%%
 
 %\cite{Mehen:2017nrh}
\bibitem{Mehen:2017nrh}
T.~Mehen,
Implications of Heavy Quark-Diquark Symmetry for Excited Doubly Heavy Baryons and Tetraquarks,
Phys. Rev. D \textbf{96}, 094028 (2017)
%doi:10.1103/PhysRevD.96.094028
[arXiv:1708.05020].
 
%\cite{Bicudo:2015vta}
\bibitem{Bicudo:2015vta} 
  P.~Bicudo, K.~Cichy, A.~Peters, B.~Wagenbach and M.~Wagner,
Evidence for the existence of $u d \bar{b} \bar{b}$ and the non-existence of $s s \bar{b} \bar{b}$ and $c c \bar{b} \bar{b}$ tetraquarks from lattice QCD,
  Phys.\ Rev.\ D {\bf 92}, 014507 (2015)
%  doi:10.1103/PhysRevD.92.014507
  [arXiv:1505.00613].
  %%CITATION = doi:10.1103/PhysRevD.92.014507;%%
  
%\cite{Bicudo:2016ooe}
\bibitem{Bicudo:2016ooe} 
  P.~Bicudo, J.~Scheunert and M.~Wagner,
Including heavy spin effects in the prediction of a $\bar{b} \bar{b} u d$ tetraquark with lattice QCD potentials,
  Phys.\ Rev.\ D {\bf 95}, 034502 (2017)
%  doi:10.1103/PhysRevD.95.034502
  [arXiv:1612.02758].
  %%CITATION = doi:10.1103/PhysRevD.95.034502;%%    
    
%\cite{Karliner:2017qjm}
\bibitem{Karliner:2017qjm} 
  M.~Karliner and J.L.~Rosner,
Discovery of doubly-charmed $\Xi_{cc}$ baryon implies a stable ($b b \bar{u} \bar{d}$) tetraquark,
  Phys.\ Rev.\ Lett.\  {\bf 119}, 202001 (2017)
%  doi:10.1103/PhysRevLett.119.202001
  [arXiv:1707.07666].
  %%CITATION = doi:10.1103/PhysRevLett.119.202001;%%
  
%\cite{Eichten:2017ffp}
\bibitem{Eichten:2017ffp} 
  E.J.~Eichten and C.~Quigg,
Heavy-quark symmetry implies stable heavy tetraquark mesons $Q_iQ_j \bar q_k \bar q_l$,
  Phys.\ Rev.\ Lett.\  {\bf 119}, 202002 (2017)
%  doi:10.1103/PhysRevLett.119.202002
  [arXiv:1707.09575].
  

%\cite{Lu:2020rog}
\bibitem{Lu:2020rog}
Q.F.~L\"u, D.Y.~Chen and Y.B.~Dong,
Masses of doubly heavy tetraquarks $T_{QQ^\prime}$ in a relativized quark model,
Phys.\ Rev.\ D \textbf{102}, 034012 (2020)
%doi:10.1103/PhysRevD.102.034012
[arXiv:2006.08087].

  
%\cite{Francis:2016hui}
\bibitem{Francis:2016hui} 
  A.~Francis, R.J.~Hudspith, R.~Lewis and K.~Maltman,
Lattice Prediction for Deeply Bound Doubly Heavy Tetraquarks,
  Phys.\ Rev.\ Lett.\  {\bf 118}, 142001 (2017)
%  doi:10.1103/PhysRevLett.118.142001
  [arXiv:1607.05214].
  %%CITATION = doi:10.1103/PhysRevLett.118.142001;%%
  
%\cite{Francis:2018jyb}
\bibitem{Francis:2018jyb} 
  A.~Francis, R.J.~Hudspith, R.~Lewis and K.~Maltman,
Evidence for charm-bottom tetraquarks and the mass dependence of heavy-light tetraquark states from lattice QCD,
  Phys.\ Rev.\ D {\bf 99}, 054505 (2019)
%  doi:10.1103/PhysRevD.99.054505
  [arXiv:1810.10550].
  %%CITATION = doi:10.1103/PhysRevD.99.054505;%%

%\cite{Cheung:2017tnt}
\bibitem{Cheung:2017tnt} 
  G.K.C.~Cheung {\it et al.} [Hadron Spectrum Collaboration],
Tetraquark operators in lattice QCD and exotic flavour states in the charm sector,
  JHEP {\bf 1711}, 033 (2017)
%  doi:10.1007/JHEP11(2017)033
  [arXiv:1709.01417].
  %%CITATION = doi:10.1007/JHEP11(2017)033;%% 
  
%\cite{Junnarkar:2018twb}
\bibitem{Junnarkar:2018twb} 
  P.~Junnarkar, N.~Mathur and M.~Padmanath,
Study of doubly heavy tetraquarks in Lattice QCD,
  Phys.\ Rev.\ D {\bf 99}, 034507 (2019)
%  doi:10.1103/PhysRevD.99.034507
  [arXiv:1810.12285].
  %%CITATION = doi:10.1103/PhysRevD.99.034507;%%

 %\cite{Leskovec:2019ioa}
\bibitem{Leskovec:2019ioa} 
  L.~Leskovec, S.~Meinel, M.~Pflaumer and M.~Wagner,
Lattice QCD investigation of a doubly-bottom $\bar{b} \bar{b} u d$ tetraquark with quantum numbers $I(J^P) = 0(1^+)$,
  Phys.\ Rev.\ D {\bf 100}, 014503 (2019)
%  doi:10.1103/PhysRevD.100.014503
  [arXiv:1904.04197].
  %%CITATION = doi:10.1103/PhysRevD.100.014503;%%
  
%\cite{Aaij:2017ueg}
\bibitem{Aaij:2017ueg} 
  R.~Aaij {\it et al.} [LHCb Collaboration],
Observation of the doubly charmed baryon $\Xi_{cc}^{++}$,
  Phys.\ Rev.\ Lett.\  {\bf 119},  112001 (2017)
%  doi:10.1103/PhysRevLett.119.112001
  [arXiv:1707.01621].
  %%CITATION = doi:10.1103/PhysRevLett.119.112001;%%
 
%\cite{Karliner:2014gca}
\bibitem{Karliner:2014gca} 
  M.~Karliner and J.L.~Rosner,
Baryons with two heavy quarks: Masses, production, decays, and detection,
  Phys.\ Rev.\ D {\bf 90}, 094007 (2014)
%  doi:10.1103/PhysRevD.90.094007
  [arXiv:1408.5877].
  %%CITATION = doi:10.1103/PhysRevD.90.094007;%% 
 
%\cite{Georgi:1990um}
\bibitem{Georgi:1990um}
H.~Georgi,
An Effective Field Theory for Heavy Quarks at Low-energies,
Phys. Lett. B \textbf{240}, 447-450 (1990).
%doi:10.1016/0370-2693(90)91128-X

%\cite{Tanabashi:2018oca}
\bibitem{Tanabashi:2018oca} 
P.A.\ Zyla et al. (Particle Data Group), to be published in Prog.\ Theor.\ Exp.\ Phys.\ {\bf 2020}, 083C01 (2020).  

  %\cite{Briceno:2012wt}
\bibitem{Briceno:2012wt} 
  R.A.~Briceno, H.W.~Lin and D.R.~Bolton,
Charmed-Baryon Spectroscopy from Lattice QCD with $N_f=2+1+1$ Flavors,
  Phys.\ Rev.\ D {\bf 86}, 094504 (2012)
%  doi:10.1103/PhysRevD.86.094504
  [arXiv:1207.3536].
  %%CITATION = doi:10.1103/PhysRevD.86.094504;%%
  
  %\cite{Alexandrou:2014sha}
\bibitem{Alexandrou:2014sha} 
  C.~Alexandrou, V.~Drach, K.~Jansen, C.~Kallidonis and G.~Koutsou,
Baryon spectrum with $N_f=2+1+1$ twisted mass fermions,
  Phys.\ Rev.\ D {\bf 90}, 074501 (2014)
%  doi:10.1103/PhysRevD.90.074501
  [arXiv:1406.4310].
  %%CITATION = doi:10.1103/PhysRevD.90.074501;%%

%\cite{Brown:2014ena}
\bibitem{Brown:2014ena} 
  Z.S.~Brown, W.~Detmold, S.~Meinel and K.~Orginos,
Charmed bottom baryon spectroscopy from lattice QCD,
  Phys.\ Rev.\ D {\bf 90}, 094507 (2014)
%  doi:10.1103/PhysRevD.90.094507
  [arXiv:1409.0497].
  %%CITATION = doi:10.1103/PhysRevD.90.094507;%%
  

%\cite{Liu:2009jc}
\bibitem{Liu:2009jc}
L.~Liu, H~W.~Lin, K.~Orginos and A.~Walker-Loud,
Singly and Doubly Charmed J=1/2 Baryon Spectrum from Lattice QCD,
Phys.\ Rev.\ D \textbf{81}, 094505 (2010)
%doi:10.1103/PhysRevD.81.094505
[arXiv:0909.3294].
  
%\cite{Namekawa:2013vu}
\bibitem{Namekawa:2013vu}
Y.~Namekawa \textit{et al.} [PACS-CS],
Charmed baryons at the physical point in 2+1 flavor lattice QCD,
Phys.\ Rev.\ D \textbf{87}, 094512 (2013)
%doi:10.1103/PhysRevD.87.094512
[arXiv:1301.4743].

%\cite{Bali:2015lka}
\bibitem{Bali:2015lka} 
  P.~P\'erez-Rubio, S.~Collins and G.S.~Bali,
Charmed baryon spectroscopy and light flavor symmetry from lattice QCD,
  Phys.\ Rev.\ D {\bf 92}, 034504 (2015)
%  doi:10.1103/PhysRevD.92.034504
  [arXiv:1503.08440].

%\cite{Alexandrou:2017xwd}
\bibitem{Alexandrou:2017xwd} 
  C.~Alexandrou and C.~Kallidonis,
Low-lying baryon masses using $N_f=2$ twisted mass clover-improved fermions directly at the physical pion mass,
  Phys.\ Rev.\ D {\bf 96}, 034511 (2017)
%  doi:10.1103/PhysRevD.96.034511
  [arXiv:1704.02647].
  %%CITATION = doi:10.1103/PhysRevD.96.034511;%%

\bibitem{Bahtiyar:2020uuj}
H.~Bahtiyar, K.U.~Can, G.~Erkol, P.~Gubler, M.~Oka and T.T.~Takahashi,
Charmed baryon spectrum from lattice QCD near the physical point,
Phys.\ Rev.\ D \textbf{102}, 054513 (2020)
%doi:10.1103/PhysRevD.102.054513
[arXiv:2004.08999].

%\cite{Lewis:2008fu}
\bibitem{Lewis:2008fu}
R.~Lewis and R.~Woloshyn,
Bottom baryons from a dynamical lattice QCD simulation,
Phys.\ Rev.\ D \textbf{79}, 014502 (2009)
%doi:10.1103/PhysRevD.79.014502
[arXiv:0806.4783].
  
%\cite{Mohanta:2019mxo}
\bibitem{Mohanta:2019mxo}
P.~Mohanta and S.~Basak,
Heavy baryon spectrum on lattice with NRQCD bottom and HISQ lighter quarks,
Phys. Rev. D \textbf{101}, 094503 (2020)
%doi:10.1103/PhysRevD.101.094503
[arXiv:1911.03741 [hep-lat]].

%\cite{Brambilla:1999xf}
\bibitem{Brambilla:1999xf}
N.~Brambilla, A.~Pineda, J.~Soto and A.~Vairo,
Potential NRQCD: An Effective theory for heavy quarkonium,
Nucl. Phys. B \textbf{566}, 275 (2000)
%doi:10.1016/S0550-3213(99)00693-8
[arXiv:hep-ph/9907240].

%\cite{Mehen:2019cxn}
\bibitem{Mehen:2019cxn}
T.C.~Mehen and A.~Mohapatra,
Perturbative Corrections to Heavy Quark-Diquark Symmetry Predictions for Doubly Heavy Baryon Hyperfine Splittings,
Phys.\ Rev.\ D \textbf{100}, 076014 (2019)
%doi:10.1103/PhysRevD.100.076014
[arXiv:1905.06965].

%\cite{Mathur:2018rwu}
\bibitem{Mathur:2018rwu} 
  N.~Mathur and M.~Padmanath,
Lattice QCD study of doubly-charmed strange baryons,
  Phys.\ Rev.\ D {\bf 99}, 031501 (2019)
%  doi:10.1103/PhysRevD.99.031501
  [arXiv:1807.00174].
  %%CITATION = doi:10.1103/PhysRevD.99.031501;%%

%\cite{Aaij:2018gfl}
\bibitem{Aaij:2018gfl}
R.~Aaij \textit{et al.} [LHCb],
First Observation of the Doubly Charmed Baryon Decay $\Xi_{cc}^{++}\rightarrow \Xi_{c}^{+}\pi^{+}$,
Phys.\ Rev.\ Lett.\ \textbf{121}, 162002 (2018)
%doi:10.1103/PhysRevLett.121.162002
[arXiv:1807.01919].

%\cite{Mattson:2002vu}
\bibitem{Mattson:2002vu} 
  M.~Mattson {\it et al.} [SELEX Collaboration],
First Observation of the Doubly Charmed Baryon $\Xi^+_{cc}$,
  Phys.\ Rev.\ Lett.\  {\bf 89}, 112001 (2002)
%  doi:10.1103/PhysRevLett.89.112001
  [hep-ex/0208014].
  %%CITATION = doi:10.1103/PhysRevLett.89.112001;%%
  
%\cite{Engelfried:2005kd}
\bibitem{Engelfried:2005kd} 
  J.~Engelfried [SELEX Collaboration],
The experimental discovery of double-charm baryons,
  Nucl.\ Phys.\ A {\bf 752}, 121 (2005).
%  doi:10.1016/j.nuclphysa.2005.02.031
  %%CITATION = doi:10.1016/j.nuclphysa.2005.02.031;%%  

%\cite{Juge:1999ie}
\bibitem{Juge:1999ie} 
  K.J.~Juge, J.~Kuti and C.J.~Morningstar,
Ab initio study of hybrid $\bar b g b$ mesons,
  Phys.\ Rev.\ Lett.\  {\bf 82}, 4400 (1999)
%  doi:10.1103/PhysRevLett.82.4400
  [hep-ph/9902336].
  %%CITATION = doi:10.1103/PhysRevLett.82.4400;%%
  
%\cite{Braaten:2014qka}
\bibitem{Braaten:2014qka} 
  E.~Braaten, C.~Langmack and D.H.~Smith,
Born-Oppenheimer approximation for the $XYZ$ mesons,
  Phys.\ Rev.\ D {\bf 90}, 014044 (2014)
%  doi:10.1103/PhysRevD.90.014044
  [arXiv:1402.0438].
  %%CITATION = doi:10.1103/PhysRevD.90.014044;%%
  
  

      
%\cite{Brambilla:2017uyf}
\bibitem{Brambilla:2017uyf}
N.~Brambilla, G.~Krein, J.J.~Tarr\'us Castell\`a and A.~Vairo,
Born-Oppenheimer approximation in an effective field theory language,
Phys.\ Rev.\ D \textbf{97}, 016016 (2018)
% doi:10.1103/PhysRevD.97.016016
[arXiv:1707.09647].

%\cite{Berwein:2015vca}
\bibitem{Berwein:2015vca}
M.~Berwein, N.~Brambilla, J.J.~Tarr\'us Castell\`a and A.~Vairo,
Quarkonium Hybrids with Nonrelativistic Effective Field Theories,
Phys.\ Rev.\ D \textbf{92}, 114019 (2015)
%doi:10.1103/PhysRevD.92.114019
[arXiv:1510.04299].

%\cite{Oncala:2017hop}
\bibitem{Oncala:2017hop}
R.~Oncala and J.~Soto,
Heavy Quarkonium Hybrids: Spectrum, Decay and Mixing,
Phys.\ Rev.\ D \textbf{96}, 014004 (2017)
% doi:10.1103/PhysRevD.96.014004
[arXiv:1702.03900].

%\cite{Brambilla:2018pyn}
\bibitem{Brambilla:2018pyn}
N.~Brambilla, W.K.~Lai, J.~Segovia, J.~Tarr\'us Castell\`a and A.~Vairo,
Spin structure of heavy-quark hybrids,
Phys.\ Rev.\ D \textbf{99}, 014017 (2019)
%doi:10.1103/PhysRevD.99.014017
[arXiv:1805.07713].

%\cite{Brambilla:2019jfi}
\bibitem{Brambilla:2019jfi}
N.~Brambilla, W.K.~Lai, J.~Segovia and J.~J.~Tarr\'us Castell\`a,
QCD spin effects in the heavy hybrid potentials and spectra,
Phys.\ Rev.\ D \textbf{101}, 054040 (2020)
%doi:10.1103/PhysRevD.101.054040
[arXiv:1908.11699].

%\cite{Bicudo:2012qt}
\bibitem{Bicudo:2012qt} 
  P.~Bicudo {\it et al.} [European Twisted Mass Collaboration],
Lattice QCD signal for a bottom-bottom tetraquark,
  Phys.\ Rev.\ D {\bf 87},  114511 (2013)
%  doi:10.1103/PhysRevD.87.114511
  [arXiv:1209.6274].
  %%CITATION = doi:10.1103/PhysRevD.87.114511;%%  

%\cite{Soto:2020xpm}
\bibitem{Soto:2020xpm}
J.~Soto and J.~Tarr\'us Castell\`a,
Nonrelativistic effective field theory for heavy exotic hadrons,
Phys.\ Rev.\ D \textbf{102}, 014012 (2020)
%doi:10.1103/PhysRevD.102.014012
[arXiv:2005.00552].

%\cite{Soto:2020pfa}
\bibitem{Soto:2020pfa}
J.~Soto and J.~Tarr\'us Castell\`a,
Effective field theory for double heavy baryons at strong coupling,
Phys.\ Rev.\ D \textbf{102}, 014013 (2020)
%doi:10.1103/PhysRevD.102.014013
[arXiv:2005.00551].
  
%%%%%%%%%%%%%%%%%%%%%%%%%%%%%%%%%%%%%%%%%%%%%%%
\end{thebibliography}
\end{document}